\documentclass[12pt]{article}

\usepackage{graphicx}
\usepackage{floatflt, graphicx}
\usepackage{amssymb}
\usepackage{subfigure}

\newcommand{\diffunit}{$\mathrm{GeV\;cm^{-2}\;s^{-1}\;sr^{-1}}$}

\newcommand{\esqdnde}{$E^{2}_{\nu} \times dN_{\nu}/dE_{\nu}$}

\begin{document}

\title {\bf Astroparticle Physics with High Energy Neutrinos: from AMANDA to IceCube}
\author{Francis Halzen\\
Department of Physics, University of Wisconsin, Madison, WI 53706}

\date{}

\maketitle

\begin{abstract}
Kilometer-scale neutrino detectors such as IceCube are discovery instruments covering nuclear and particle physics, cosmology and astronomy. Examples of their multidisciplinary missions include the search for the particle nature of dark matter and for additional small dimensions of space. In the end, their conceptual design is very much anchored to the observational fact that Nature produces protons and photons with energies in excess of $10^{20}$\,eV and $10^{13}$\,eV, respectively. The puzzle of where and how Nature accelerates the highest energy cosmic particles is unresolved almost a century after their discovery. The cosmic ray connection sets the scale of cosmic neutrino fluxes. In this context, we discuss the first results of the completed AMANDA detector and the science reach of its extension, IceCube. Similar experiments are under construction in the Mediterranean. Neutrino astronomy is also expanding in new directions with efforts to detect air showers, acoustic and radio signals initiated by super-EeV neutrinos. The outline of this review is as follows:
\begin{itemize}
\item Introduction
\item Why Kilometer-Scale Detectors?
\item Cosmic Neutrinos Associated with the Highest Energy Cosmic Rays
\item High Energy Neutrino Telescopes: Methodologies of Neutrino Detection
\item High Energy Neutrino Telescopes: Status
\end{itemize}
\end{abstract}

\section{Introduction}

Ambitious projects have been launched to extend conventional astronomy beyond wavelengths of $10^{-14}$\,cm, or GeV photon energy. Besides gamma rays, protons (nuclei), neutrinos and gravitational waves will be explored as astronomical messengers probing the extreme Universe. The challenges are considerable:
\begin{itemize}
\item Protons are relatively abundant, but their arrival directions have been scrambled by magnetic fields.
\item $\gamma$-rays do point back to their sources, but are absorbed at TeV-energy and above on cosmic background radiation.
\item neutrinos propagate unabsorbed and without deflection throughout the universe but are difficult to detect.
\end{itemize}
Therefore, multi-messenger astronomy may not just be an advantage, it may be a necessity for solving some of the outstanding problems of astronomy at the highest energies such as the identification of the sources of the cosmic rays, the mechanism(s) triggering gamma ray bursts and the particle nature of the dark matter.

We will update the case for the detection of neutrinos associated with the observed fluxes of high energy cosmic rays and gamma rays; it points, unfortunately, at the necessity of commissioning kilometer-scale neutrino detectors. Though ambitious, the scientific case is compelling because neutrinos will reveal the location of the source(s) and represent the ideal tool to study the black holes powering the cosmic accelerator(s).

At this time the first-generation neutrino telescope AMANDA has operated for 5 years and a second one, ANTARES, is under construction in the Mediterranean. AMANDA represents a proof of concept for a kilometer-scale detector, IceCube, now under construction. High energy neutrino astronomy predates these projects; we have observed the Sun and a supernova in 1987\cite{totsuka}. Each observation has been rewarded with a Nobel Prize. These achievements were influential. After thirty years the solar neutrino puzzle was resolved by the discovery that neutrinos oscillate. The skeptics were proven wrong, John Bahcall knew all along how the Sun shines. Some 20 supernova neutrinos were adequate to confirm the basic theoretical picture of the death of a star. The goal of neutrino telescopes is to look beyond the Sun, possibly to the edge of the Universe. Construction of IceCube and other high-energy neutrino telescopes is mostly motivated by their potential to open a new window on the Universe using neutrinos as cosmic messengers. This will be the central topic of this review.

Soon after the discovery in the mid-fifties that neutrinos where real particles and not just mathematical constructs of theorists' imagination, the idea emerged that they represent ideal cosmic messengers\cite{reines}. Because of their weak interactions, neutrinos reach us unimpeded from the edge of the Universe and from the inner reaches of black holes. The neutrino telescopes now under construction have the capability to detect neutrinos with energies from a threshold of $\sim 10$\,GeV to, possibly, $ \sim 10^2$\,EeV, the highest energies observed. Their telescope range spans more than 10 orders of magnitude in wavelengths smaller than $10^{-14}$\,cm. This is a reach equivalent to that of a hypothetical astronomical telescope sensitive to wavelengths from radio to X-rays. Above $10^5$\,TeV the observations are free of muon and neutrino backgrounds produced in cosmic ray interactions with the Earth's atmosphere. Each neutrino is a discovery.\footnote{We will use  GeV$=10^9$\,eV, TeV$=10^{12}$\,eV,  PeV$=10^{15}$\,eV and EeV$=10^{18}$\,eV units of energy.}

The real challenge of neutrino astronomy is that kilometer-scale neutrino detectors are required to do the science. The first hint of the scale of neutrino telescopes emerged in the nineteen seventies from theoretical studies of the flux of neutrinos produced in the interactions of cosmic rays with microwave photons, the so-called Greissen-Zatsepin-Kuzmin or GZK neutrinos. Since then the case for kilometer-size instruments has been strengthened\cite{PR} and the possibility of commissioning such instruments demonstrated\cite{ice3}. In fact, if the neutrino sky were within reach of smaller instruments, it would by now have been revealed by the first-generation AMANDA telescope. It has been taking data since 2000 with a detector of $0.01 \sim 0.08$\,km$^2$ telescope area, depending on the sources\cite{pune}.

A wealth of particle physics has been extracted from a small sample of supernova neutrinos and the most exciting result of solar neutrino astronomy to date is not related to the Sun but to the neutrino itself. With the discovery of neutrino mass in underground experiments, particle astrophysics has indeed reconnected with the early cosmic ray tradition of doing fundamental particle physics with heavenly beams. History may repeat itself with the advent of high energy neutrino telescopes\cite{GHM}. As for conventional astronomy, we have to observe the neutrino sky through the atmosphere. This is a curse and a blessing; the background of neutrinos produced by cosmic rays in interactions with atmospheric nuclei also provides a beam for calibration of the experiments. It also presents us with an opportunity to do particle physics\cite{GHM}. Especially unique is the energy range covering $0.1 \sim 10^5$~TeV of the ``background" atmospheric neutrinos accumulated by neutrino ``telescopes" and definitely not within reach of accelerators. Cosmic beams of even higher energy may exist, but the atmospheric beam is guaranteed.

Construction of IceCube and other high-energy neutrino telescopes is
mostly motivated by their potential to open a new window on the
Universe using neutrinos as cosmic messengers. The IceCube experiment
nevertheless appeared on the U.S.\ Roadmap to Particle
Physics\cite{baggerbarish}, and deservedly so. As
the lightest of fermions and the most weakly interacting of particles,
neutrinos occupy a fragile corner of the Standard Model and one can
realistically hope that they will reveal the first and most dramatic
signatures of new physics.

IceCube's opportunities for particle physics are only limited by
imagination\cite{icehep}; they include:
\begin{enumerate}
  \item The search for neutrinos from the annihilation of dark matter
    particles gravitationally trapped at the center of the Sun and the
    Earth\cite{bertone}.
  \item The search for the signatures of the possible unification of particle interactions, including
    gravity, at the TeV scale.  Neutrinos with energies
    approaching this scale would interact by gravity with large
    cross sections, similar to those of quarks and leptons, and this
    increase should yield dramatic signatures in a neutrino telescope
    including, possibly, the production of black holes\cite{feng}.
  \item The search for deviations from the neutrino's established
    oscillatory behavior that result from non-standard neutrino interactions.
    \item Searching for flavor changes of neutrino beams over cosmic distances as a signature for quantum decoherence.
  \item The search for a breakdown of the equivalence principle as a
    result of non-universal interactions with the gravitational field
    of neutrinos with different flavor.
  \item Similarly, the search for breakdown of Lorentz invariance
    resulting from different limiting velocities of neutrinos of
    different flavors.  With energies of $10^3$~TeV and masses of order $10^{-2}$~eV or less, even the atmospheric neutrinos observed by IceCube reach Lorentz factors of $10^{17}$ or larger.
  \item The search for particle emission from cosmic strings or any other
    topological defects or heavy cosmological remnants created in the
    early Universe. It has been suggested that they may be the sources of the highest energy cosmic rays.
  \item The search for magnetic monopoles, Q-balls and the like.
\end{enumerate}

It is well-known that oscillations are not the only mechanism for
atmospheric $\nu_\mu \to \nu_\tau$ flavour
transitions\cite{npreview}. These can result from non-standard neutrino interactions 
that mix neutrino flavours.  Examples include violations of the
equivalence principle, non-standard
neutrino interactions with matter, neutrino couplings to
space-time torsion fields, violations of Lorentz
invariance and of CPT
symmetry.  Although these scenarios no
longer explain the existing data\cite{oldatmfitnp}, a combined
analysis of the atmospheric neutrino and K2K data can be performed to
obtain the best constraints to date on the size of such subdominant
oscillation effects\cite{ouratmnp}.

A critical feature of new-physics scenarios is that they introduce a
departure from the characteristic energy $L/E$ dependence associated with
the mass-induced oscillation wavelength $L$.  New physics introduces modifications that are constant or increase with energy and therefore IceCube, with
energy reach of $0.1 \sim 10^5$~TeV for atmospheric
neutrinos, will have unmatched sensitivity. Furthermore, in most of this energy interval conventional oscillations are
suppressed and therefore the observation of an angular distortion of
the atmospheric neutrino flux or its energy dependence will provide
signatures for the presence of new physics mixing neutrino flavors
that are not obscured by oscillations associated with their mass.

IceCube is expected to collect a data set of order one million neutrinos over 10 years. Not surprisingly, because of the increased energy and statistics
over present experiments, sensitivity to violations of the
equivalence principle and of Lorentz invariance, for instance, will
be improved by over two orders of magnitude; see\cite{GHM}.

Given the size of the detector required, all efforts have concentrated on transforming large volumes of natural water or  ice into Cherenkov detectors. They reveal the secondary muons and electromagnetic and hadronic showers initiated in neutrino interactions inside or near the detector. Because of the long range of the muon, from kilometers in the TeV range to tens of kilometers at the highest energies, neutrino interactions can be identified far outside the instrumented volume. Adding to the technological challenge is the requirement that the detector be shielded from the abundant flux of cosmic ray muons by deployment at a depth of typically several kilometers. After the cancellation of a pioneering attempt\cite{water} to build a neutrino telescope off the coast of Hawaii, successful operation of a smaller instrument in Lake Baikal\cite{baikal} bodes well for several efforts to commission neutrino telescopes in the Mediterranean\cite{water,emigneco}. We will here mostly concentrate on the construction and first four years of operation of the AMANDA telescope\cite{pune,nature} which has transformed a large volume of natural deep Antarctic ice into a Cherenkov detector. It represents a first-generation telescope as envisaged by the DUMAND collaboration over 20 years ago and a proof of concept for the kilometer-scale IceCube detector, now under construction.

Even though neutrino ``telescopes" are designed as discovery instruments covering a large dynamic range, be it for particle physics or astrophysics, their conceptual design is very much anchored to the observational fact that Nature produces protons and photons with energies in excess of $10^{20}$\.eV and $10^{13}$\,eV, respectively. The cosmic ray connection sets the scale of cosmic neutrino fluxes. We will discuss this first.

\section{The Scale of Neutrino Telescopes:\\ Cosmic Neutrinos associated with\\ the Highest Energy Cosmic Rays}
 
Cosmic accelerators produce particles with energies in excess of $10^8$\,TeV; we do not know where or how. The flux of cosmic rays observed at Earth is sketched in Fig.\,1a,b\cite{gaisseramsterdam}. The  energy spectrum follows a broken power law. The two power laws are separated by a feature dubbed the ``knee"; see Fig.\,1a. Circumstantial evidence exists that cosmic rays, up to perhaps EeV energy, originate in galactic supernova remnants. Any association with our Galaxy disappears in the vicinity of a second feature in the spectrum referred to as the ``ankle". Above the ankle, the gyroradius of a proton in the galactic magnetic field exceeds the size of the Galaxy and it is generally assumed that we are  witnessing the onset of an extragalactic component in the spectrum that extends to energies beyond 100\,EeV. Experiments indicate that the highest energy cosmic rays are predominantly protons or, possibly, nuclei. Above a threshold of 50 EeV these protons interact with cosmic microwave photons and lose energy to pions before reaching our detectors. This is the GZK cutoff that limits the sources to our local supercluster. 

\begin{figure}[h]
\centering\leavevmode
\includegraphics[width=5.8in]{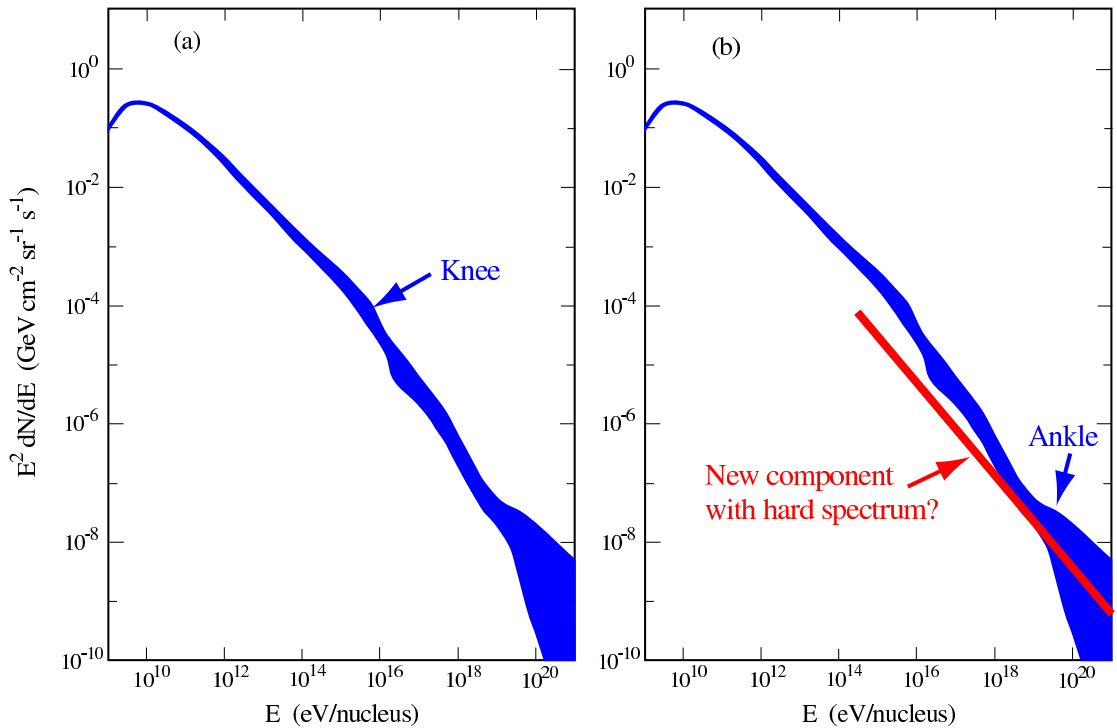}
\caption{At the energies of interest here, the cosmic ray spectrum consists of a sequence of 3 power laws. The first two are separated by the ``knee" (left panel), the second and third by the "ankle". There is evidence that the cosmic rays beyond the ankle are a new population of particles produced in extragalactic sources; see right panel.}
\label{knee-ankle}
\end{figure}

Models for the origin of the highest energy cosmic rays fall into two categories, top-down and bottom-up. In top-down models it is assumed that the cosmic rays are the decay products of cosmological remnants or topological defects associated, for instance, with Grand Unified theories with unification energy $M_{GUT} \sim 10^{24}\rm\,eV$. These models predict neutrino fluxes most likely within reach of first-generation telescopes such as AMANDA, and certainly detectable by future kilometer-scale neutrino observatories\cite{semikoz}. They have not been observed.

In bottom-up scenarios it is assumed that cosmic rays originate in cosmic accelerators. Accelerating particles to TeV energy and above requires massive bulk flows of relativistic charged particles. These are likely to originate from the exceptional gravitational forces  in the vicinity of black holes. Gravity powers large electric currents that create the opportunity for particle acceleration by shocks, a mechanism familiar from solar flares where particles are accelerated to $10$\,GeV. It is a fact that black holes accelerate electrons to high energy; astronomers observe them indirectly by their synchrotron radiation. We know that they accelerate protons because we detect them as cosmic rays. Because they are charged, protons are deflected by interstellar magnetic fields; cosmic rays do not reveal their sources. This is the cosmic ray puzzle.

Examples of candidate black holes include the dense cores of exploding stars, inflows onto supermassive black holes at the centers of active galaxies and annihilating black holes or neutron stars. Before leaving the source, accelerated particles pass through intense radiation fields or dense clouds of gas surrounding the black hole. This results in interactions producing pions decaying into secondary photons and neutrinos that accompany the primary cosmic ray beam as illustrated in Fig.\,2.
%
\begin{figure}[!h]
\centering\leavevmode
\includegraphics[width=4.25in]{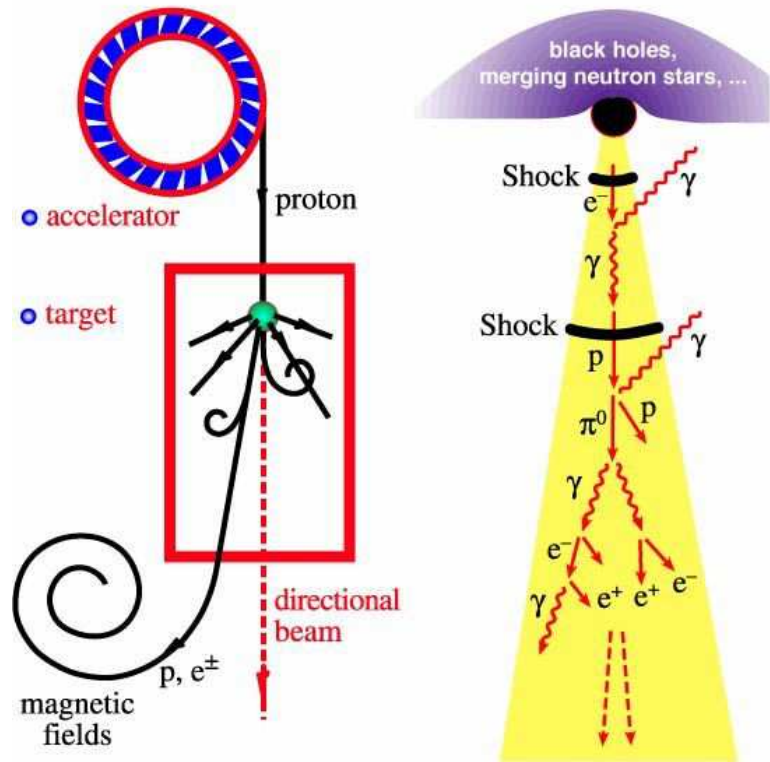}
\caption{Cosmic beam dump exits: sketch of cosmic ray accelerator producing photons. The charged pions that are inevitably produced along with the neutral pions will decay into neutrinos.}
\label{nubeams}
\end{figure}
How many neutrinos are produced in association with the cosmic ray beam? The answer to this question, among many others\cite{PR},  provides the rationale for building kilometer-scale neutrino detectors.

First consider a neutrino beam produced at an accelerator laboratory; see Fig.\,2. Here the target absorbs all parent protons as well as the secondary electromagnetic and hadronic showers. Only neutrinos exit the dump. If Nature constructed such a "hidden source" in the heavens, conventional astronomy will not reveal it. It cannot be the source of the cosmic rays, however, because in this case the dump must be transparent to protons. A more generic "transparent" source can be envisaged as follows: protons are accelerated near a black hole in regions of high magnetic fields resulting from shocks. They interact with photons surrounding the accelerator via the processes 
%
$p + \gamma \rightarrow \Delta \rightarrow \pi^0 + p$
and
$p + \gamma \rightarrow \Delta \rightarrow \pi^+ + n$.
%
While the secondary protons may remain trapped in the acceleration region, equal numbers of neutrons, neutral and charged pions escape. The energy escaping the source is therefore distributed between cosmic rays, gamma rays and neutrinos produced by the decay of neutrons and neutral and charged pions, respectively. The neutrino flux from a generic transparent cosmic ray source is often referred to as the Waxman-Bahcall flux\cite{wb1}. It is easy to calculate and the derivation is revealing.

Figure 1b shows a fit to the observed spectrum above the ``ankle" that can be used to derive the total energy in extragalactic cosmic rays. The flux above the ankle is often summarized as ``one $10^{19}$\,eV particle per kilometer square per year per steradian". This can be translated into an energy flux
\begin{eqnarray*}
E \left\{ E{dN\over dE} \right\} = {10^{19}\,{\rm eV} \over \rm (10^{10}\,cm^2)(3\times 10^7\,sec) \, sr}
= 3\times 10^{-8}\rm\, GeV\ cm^{-2} \, s^{-1} \, sr^{-1} \,.
\end{eqnarray*}
From this we can derive the energy density $\rho_E$ in cosmic rays using the relation that flux${}={}$velocity${}\times{}$density, or
\[
4\pi \int  dE \left\{ E{dN\over dE} \right\} =  c\rho_E\,.
\]
We obtain
\[
\rho_e = {4\pi\over c} \int_{E_{\rm min}}^{E_{\rm max}} {3\times 10^{-8}\over E} dE \, {\rm {GeV\over cm^3}} \simeq 10^{-19} \, {\rm {TeV\over cm^3}} \,,
\]
taking the extreme energies of the accelerator(s) to be $E_{\rm max} / E_{\rm min} \simeq 10^3$.

The energy content derived ``professionally" by integrating the spectrum in Fig.~1b assuming an $E^{-2}$ energy spectrum, typical of shock acceleration, with a GZK cutoff  is $\sim 3 \times 10^{-19}\rm\,erg\ cm^{-3}$. This is within a factor of our back-of-the-envelope estimate (1\,TeV = 1.6\,erg). The power required for a population of sources to generate this energy density over the Hubble time of $10^{10}$\,years is $\sim 3 \times 10^{37}\rm\,erg\ s^{-1}$ per (Mpc)$^3$ or, as often quoted in the literature, $\sim 5\times10^{44}\rm\,TeV$ per (Mpc)$^3$ per year. This works out to\cite{TKG}
\begin{itemize}
\item $\sim 3 \times 10^{39}\rm\,erg\ s^{-1}$ per galaxy,
\item $\sim 3 \times 10^{42}\rm\,erg\ s^{-1}$ per cluster of galaxies,
\item $\sim 2 \times 10^{44}\rm\,erg\ s^{-1}$ per active galaxy, or
\item $\sim 2 \times 10^{52}$\,erg per cosmological gamma ray burst.
\end{itemize}
The coincidence between these numbers and the observed output in electromagnetic energy of these sources explains why they have emerged as the leading candidates for the cosmic ray accelerators. The coincidence is consistent with the relationship between cosmic rays and photons built into the "transparent" source. In the photoproduction processes roughly equal energy goes into the secondary neutrons, neutral and charged pions whose energy ends up in cosmic rays, gamma rays and neutrinos, respectively.

We therefore conclude that the same energy density  of $\rho_E \sim 3 \times 10^{-19}\rm\,erg\
 cm^{-3}$, observed in cosmic rays and electromagnetic energy, ends up in neutrinos with a spectrum $E_\nu dN / dE_{\nu}  \sim E^{-\gamma}\rm\, cm^{-2}\, s^{-1}\, sr^{-1}$ that continues up to a maximum energy $E_{\rm max}$. The neutrino flux follows from the relation
%
$ \int E_\nu dN / dE_{\nu}  =  c \rho_E / 4\pi  $.
%
For $\gamma = 1$ and $E_{\rm max} = 10^8$\,GeV, the generic source of the highest energy cosmic rays produces a flux of $ {E_\nu}^2 dN / dE_{\nu}  \sim 5 \times 10^{-8}\rm\, GeV \,cm^{-2}\, s^{-1}\, sr^{-1} $.

There are several ways to sharpen  this qualitative prediction:
\begin{itemize} 
\item The derivation fails to take into account that there are more
  cosmic rays in the Universe than observed at Earth because
  of the GZK-effect and it also neglects the evolution of the sources with redshift. This
  increases the neutrino flux, which we normalized to the observed spectrum only, by a factor $d_H/d_{\rm CMB}$, the ratio of the Hubble radius to the average attenuation length of the cosmic rays propagating in the cosmic microwave background.
\item For proton-$\gamma$ interactions muon neutrinos (and antineutrinos) receive only 1/2 of the energy of the charged pion in
  the decay chain $\pi^+\rightarrow \mu^+ +\nu_{\mu}\rightarrow e^+
  +\nu_e +\bar{\nu}_{\mu} +\nu_{\mu}$ assuming that the energy is
  equally shared between the 4 leptons. Furthermore half the muon
  neutrinos oscillate into tau neutrinos over cosmic distances.
\end{itemize}
In summary,
\begin{equation}
E_\nu{dN_\nu\over dE_\nu} = {1\over2} \times {1\over2} 
\times E{dN_{\rm CR}\over dE} \times {d_H\over d_{\rm CMB}} 
\simeq E{dN_{\rm CR}\over dE}
\end{equation}
In practise, the corrections approximately cancel. The transition from galactic to
extragalactic sources is debated; a transition at lower energy
significantly increases the energy in the extragalactic component.
This raises the possibility of an increase in the associated neutrino
flux~\cite{ringwald}.

Waxman and Bahcall referred to their flux as a bound because in
reality more energy is transferred to the neutron than to the charged
pion in the source, in the case of the photoproduction reaction $p +
\gamma \rightarrow \Delta \rightarrow \pi^+ + n$ four times more.
Therefore
\begin{equation}
E_\nu{dN_\nu\over dE_\nu} = {1\over4} E{dN_{\rm CR}\over dE} \, .
\end{equation}
In the end we estimate that the muon-neutrino flux associated with the sources of the highest energy cosmic rays is loosely confined to the range $ {E_\nu}^2 dN / dE_{\nu}= 1\sim 5 \times 10^{-8}\rm\, GeV \,cm^{-2}\, s^{-1}\, sr^{-1} $. 

The anticipated neutrino flux thus obtained has to be compared with the sensitivity of $8.9 \times 10^{-8}\rm\, GeV\ cm^{-2}\, s^{-1}\,sr^{-1}$ reached after the first 4 years of operation of the completed AMANDA detector in 2000--2003\cite{pune}. The analysis of the data has not been completed, but a limit of  $2 \times 10^{-7}\rm\,GeV\ cm^{-2}\,s^{-1}\,sr^{-1}$ has been obtained with a single year of data\cite{HS}. On the other hand, after three years of operation IceCube will reach a diffuse flux limit of $E_{\nu}^2 dN / dE_{\nu} = 2\,{\sim}\, 7 \times 10^{-9}\rm\,GeV \,cm^{-2}\, s^{-1}\, sr^{-1}$; see Fig.\,3. The exact value of the IceCube sensitivity depends on the magnitude of the dominant high energy neutrino background from the prompt decay of atmospheric charmed particles\cite{ice3}. The level of this background is difficult to anticipate theoretically and no accelerator data is available in the energy and Feynman-$ x$ range of interest\cite{ingelman}.

\begin{figure}[t]
\centering\leavevmode
\includegraphics[width=4.5in]{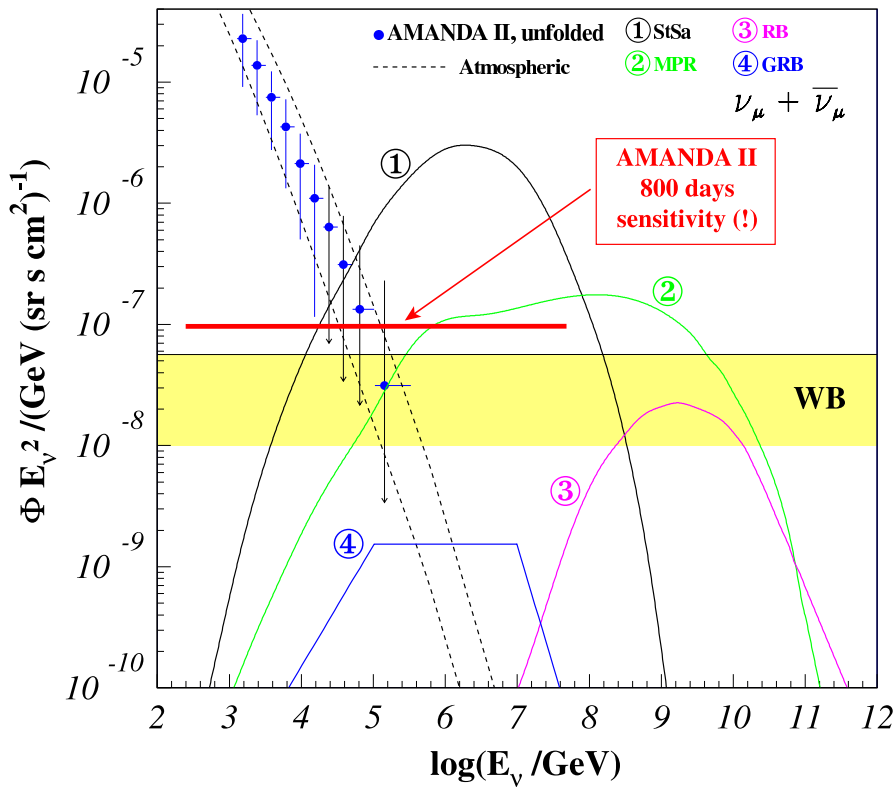}
\caption{Our estimate of the flux of neutrinos associated with the sources of the highest energy cosmic rays (the shaded range labeled WB) is compared to the sensitivity of the AMANDA experiment reached with 800 days of data. Also shown are fluxes predicted by specific models of cosmic ray accelerators: active galaxies labeled StSa\protect\cite{agn} and MPR\protect\cite{MPR}, gamma ray bursts\protect\cite{guetta} and the diffuse flux produced by cosmic ray producing active galaxies on microwave photons\protect\cite{RB} labeled RB. Data for the background atmospheric neutrino flux are from the AMANDA experiment.}
\label{fig:diffuse-incl-osc.eps}
\end{figure}

The observed event rate is obtained by folding the cosmic flux predicted with the probability that the neutrino is actually detected in a high energy neutrino telescope; only one in a million neutrinos of TeV energy interact and produce a muon that reaches the detector. This probability is given by the ratio of the muon and neutrino interaction lengths in the detector medium, $\lambda_\mu / \lambda_\nu$\cite{PR} and therefore depends on energy; this will be explained in the section on methodologies. For the flux range anticipated above we anticipate $20 \,{\sim}\, 100$ detected muon neutrinos per km$^2$ per year. Given that its effective area for muon neutrinos exceeds 1\,km$^2$ and that equal fluxes of electron and tau neutrinos are expected, a neutrino signal at the ``Waxman-Bahcall" level will result in the observation of about one thousand neutrinos in IceCube\cite{ice3}. Model calculations assuming that active galaxies or gamma-ray bursts are the actual sources of cosmic rays yield similar, or even smaller, event rates than the generic energetics estimate presented.

Gamma ray bursts (GRB), outshining the entire Universe for the duration of the burst, are perhaps the best motivated sources of high-energy neutrinos\cite{waxmanbahcall, mostlum2, mostlum3}. The collapse of massive stars to a black hole has emerged as the likely origin of the "long" GRB with durations of tens of seconds. In the collapse a fireball is produced which expands with a highly relativistic velocity powered by radiation pressure. The fireball eventually runs into the stellar material that is still accreting onto the black hole. If it successfully punctures through this stellar envelope the fireball emerges to produce a GRB. While the energy transferred to highly relativistic electrons is thus observed in the form of radiation, it is a matter of speculation how much energy is transferred to protons.

The assumption that GRB are the sources of the highest energy cosmic rays does determine the energy of the fireball baryons. Accommodating the observed cosmic ray spectrum of extragalactic cosmic rays requires roughly equal efficiency for conversion of fireball energy into the kinetic energy of protons and electrons. In this scenario the production of $100 \sim 1000$\,TeV neutrinos in the GRB fireball is a robust prediction because neutrinos are inevitably produced in interactions of accelerated protons with fireball photons. Estimates of the flux\cite{guetta} point again at the necessity of a kilometer-cubed neutrino detector, in agreement with the generic energetics estimates previously presented. Studies of active galaxies as sources of cosmic rays lead to similar conclusions\cite{agn}.

The case for kilometer-scale detectors also emerges from consideration of  ``guaranteed'' cosmic fluxes. Neutrino fluxes are guaranteed when both the accelerator and the pion producing target material can be identified. We mention three examples. The extragalactic cosmic rays produce $ \sim$ 1 event per km$^2$\,year in interactions with cosmic microwave photons\cite{cos}.  Supernovae producing  cosmic rays in the dense star forming regions of starburst galaxies form a hidden source of neutrinos within reach of IceCube\cite{loeb&wax}. Galactic cosmic rays interact with hydrogen in the disk to generate an observable neutrino flux in a kilometer-scale detector\cite{plane}.

Finally, with recent observations\cite{hess} of the supernova remnant RX J1713.7-3946 using the H.E.S.S. atmospheric Cherenkov telescope array, gamma-ray astronomy may have revealed a guaranteed source of cosmic neutrinos\cite{alvarezhalzen}. Supernova remnants have been pinpointed all along as the likely sources of the galactic cosmic rays and, with RX J1713.7-3946, H.E.S.S.  may have identified the first such site where protons are accelerated to energies typical of the main component of the galactic cosmic rays. Although the resolved image of the source (the first ever at TeV energies!) reveals TeV emission from the whole supernova remnant, it shows a clear increase of the flux in the directions of known molecular clouds. This suggests the possibility that protons, shock accelerated in the supernova remnant, interact with the dense clouds to produce neutral pions that are the source of the observed increase of the TeV photon signal. Furthermore, the high statistics data for the flux are power-law behaved over a large range of energies without any indication of a cutoff characteristic of synchrotron or inverse-Compton sources. Finally, follow-up observations of the source in radio-waves and X-rays have failed to identify the population of electrons required to generate TeV photons by purely electromagnetic processes; for a detailed discussion see \cite{hiraga}. Other interpretations are not ruled out\cite{hiraga} but, fortunately, higher statistics data is forthcoming.

If future data confirm that a fraction of the TeV flux of RX J1713.7-3946 is of neutral pion origin, then the accompanying charged pions will produce a guaranteed neutrino flux of roughly 20 muon-type neutrinos per kilometer-squared per year\cite{alvarezhalzen}. From a variety of such sources we can therefore expect event rates of cosmic neutrinos of galactic origin similar to those estimated for extragalactic neutrinos in the previous section. Supernovae associated with molecular clouds are a common feature of associations of OB stars
 that exist throughout the galactic plane.

It is important to realize that the relation between the neutrino and gamma flux is robust\cite{alvarezhalzen}. The $\nu_\mu + \bar\nu_\mu$ neutrino flux ($dN_\nu/dE_\nu$) produced by the decay of charged pions in the source can be derived from the observed gamma ray flux by energy conservation:
\begin{equation}
\int_{E_{\gamma}^{\rm min}}^{E_{\gamma}^{\rm max}}
E_\gamma {dN_\gamma\over dE_\gamma} dE_\gamma = K
\int_{E_{\nu}^{\rm min}}^{E_{\nu}^{\rm max}} E_\nu {dN_\nu\over dE_\nu} dE_\nu
\label{conservation}
\end{equation}
where ${E_{\gamma}^{\rm min}}$ ($E_{\gamma}^{\rm max}$) is the minimum (maximum) energy of the photons that have a hadronic origin. ${E_{\nu}^{\rm min}}$ and ${E_{\nu}^{\rm max}}$ are the corresponding minimum and maximum energy of the neutrinos.
The factor $K$ depends on whether the $\pi^0$'s are of $pp$ or $p\gamma$ origin. Its value can be obtained from routine particle physics. In $pp$ interactions 1/3 of the proton energy goes into each pion flavor. In the pion-to-muon-to-electron decay chain 2 muon-neutrinos are produced with energy $E_\pi/4$ for every photon with energy $E_\pi/2$. Therefore the energy in neutrinos matches the energy in photons and $K=1$. This flux has to be reduced by a factor 2 because of oscillations. For $p\gamma$ interactions $K=1/4$. The estimate should be considered a lower limit because the observed photon flux to which the calculation is normalized may have been attenuated by absorption in the source or in the interstellar medium. 

The case for doing neutrino astronomy is compelling, the challenge has been to deliver the technology to build neutrino detector that are unfortunately required to have kilometer-scale dimensions. We discuss this next.

\section{High Energy Neutrino Telescopes:\\ Methodologies}

The construction of neutrino telescopes is motivated by discovery.  To maximize this potential, one must design an instrument with the largest possible effective telescope area to overcome the small neutrino cross section with matter, and the best possible energy and angular resolution to address the wide diversity of possible signals. While the smaller first-generation detectors have been optimized to detect secondary muons initiated by $\nu_{\mu}$, kilometer-scale neutrino observatories will detect neutrinos of all flavors over a wide range of energies.

We will review the methods by which we detect neutrinos, measure their energy and identify their flavor.

\subsection{Detection Techniques}

High energy neutrinos are detected by observing the Cherenkov radiation from secondary particles produced in neutrino interactions inside large volumes of highly transparent ice or water instrumented with a lattice of photomultiplier tubes (PMT). For simplicity, assume an instrumented cubic volume of side $L$; see Fig.\,4. Also assume that the neutrino direction is perpendicular to a side of the cube. 

\begin{figure}[!h]
\centering\leavevmode
\includegraphics[width=3in]{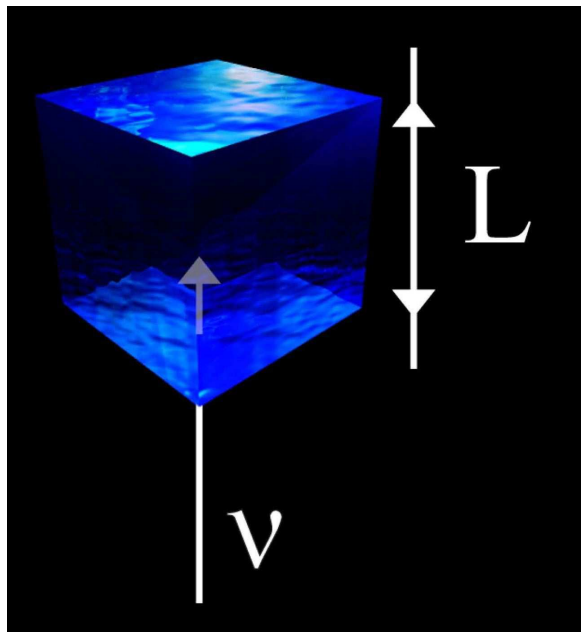}
\caption{A neutrino interacts in a cube of instrumented ice of side $L$.}
\label{fig:cube-L}
\end{figure}

To a first approximation, a neutrino of energy $E_\nu$ incident on a side of area $L^2$ will be detected provided it interacts within the detector volume, i.e.\ within the instrumented distance $L$. That probability is
\begin{equation}
P (E_\nu) = 1 - \exp[-L/\lambda_{\nu} (E_\nu)] \simeq L/\lambda_{\nu}(E_\nu) \,,
\end{equation}
where $\lambda_{\nu} (E_\nu) = [\rho_{\rm ice} \ N_A \ 
\sigma_{\nu N} (E_\nu) ]^{-1}$ is the neutrino mean free path. Here $\rho_{\rm ice} = 0.9~{\rm g}\ {\rm cm}^{-3}$ is
the density of the ice, $N_A = 6.022 \times 10^{23}$ 
is Avogadro's number and $\sigma_{\nu N}(E_\nu)$ is the neutrino-nucleon cross section.
A neutrino flux $dN/dE_\nu$ (neutrinos per GeV per cm$^2$ per s) crossing a
detector with energy threshold $E_\nu^{\rm th}$ and cross sectional area 
$A\ (= L^2)$ facing the incident beam will produce
\begin{equation}
N_{ev} = T \, \int_{E_\nu^{\rm th}} 
A(E_\nu) \, P (E_\nu) \, \frac{dN}{dE_\nu}\,\, dE_\nu
\end{equation}
events after a time $T$. In practice, the ``effective" detector area
$A$ is not strictly equal to the geometric cross section of the
instrumented volume facing the incoming neutrino because even
neutrinos interacting outside the instrumented volume may produce a
sufficient amount of light inside the detector to be detected.
Therefore, $A$ is determined as a function of the incident neutrino
direction by simulation of the full detector, including the trigger; see Appendix A.

The formalism presented applies to electron neutrinos. in the case of muon neutrinos, any neutrino producing a secondary muon that reaches the detector (and has sufficient energy to trigger it) will be detected. Because the muon travels kilometers at TeV energies and tens of kilometers at EeV energy, neutrinos can be detected outside the instrumented volume; the probability is obtained by substitution in Eq.\,4
\begin{equation}
L \rightarrow \lambda_{\mu} \,,
\end{equation}
therefore,
\begin{equation}
P = \lambda_{\mu}/\lambda_{\nu} \,.
\end{equation}
Here $ \lambda_{\mu}$ is the range of the muon determined by its energy losses. The complete expression for the flux of $\nu_\mu$-induced muons at the detector is given by a convolution of the neutrino spectrum $\phi$ (= dN/dE$_\nu$) with the probability $P$ to produce a muon reaching the detector\cite{PR}:
\begin{equation}
\phi_\mu(E_\mu^{\rm min},\theta) =
\int_{E_\mu^{\rm min}}\,P(E_\nu,E_\mu^{\rm min})\,
\exp[-\sigma_{\rm tot}(E_\nu)\,N_A\,X(\theta)]\,\phi(E_\nu,\theta).
\label{N_mu}
\end{equation}
The additional exponential factor accounts for the absorption of neutrinos along the chord of the Earth of length $X(\theta)$ at zenith angle $\theta$. Absorption becomes important for $\sigma_\nu(E_\nu)\gtrsim 10^{-33}$\,cm$^2$ or $E_\nu\gtrsim 10^7$~GeV. For back-of-the-envelope calculations, the $P$-function can be approximated by
\begin{eqnarray}
P \simeq 1.3 \times 10^{-6} E^{2.2}  && \mbox{for $E = 10^{-3}$--1 TeV} \,, \\
    \simeq 1.3 \times 10^{-6} E^{0.8}  && \mbox{for $E= 1$--$10^3$ TeV} \,.
\end{eqnarray}
At EeV energy the increase is reduced to only $E^{0.4}$. Clearly, high energy neutrinos are more likely to be detected because of the increase with energy of both the cross section and muon range; see Fig.\,5. The event rate can be calculated as for Eq.\,5; see Appendix A for details.

As an example we estimate the number of muon tracks initiated by neutrinos produced by a source at the Waxman-Bahcall level in a 1 km$^2$ during one year to be
\begin{eqnarray}
N & = &{ \rm Area \times time \times 2\pi} \, \int_{E_\mu^{\rm min}}\,P(E_\nu,E_\mu^{\rm min})\,\phi(E_\nu,\theta)  \nonumber \\
& \sim & 10^{10} \times 3 \times 10^7 \times 2\pi \int_{E_\mu^{\rm min}}\, [10^{-9} E] [3\times 10^{-8}\rm\, GeV\ cm^{-2} \, s^{-1} \, sr^{-1}] \nonumber \\
& \sim & 100 \times \ln \left[\frac {E_\mu^{\rm max}}{E_\mu^{\rm min}} \right]\,,
\label{N_mu}
\end{eqnarray}
where the logarithm depends on $E_\mu^{\rm max}$ the maximum energy of the cosmic accelerator; we used GeV units.
\begin{figure}[t]
\begin{center}
\subfigure{\includegraphics[width=3.1in]{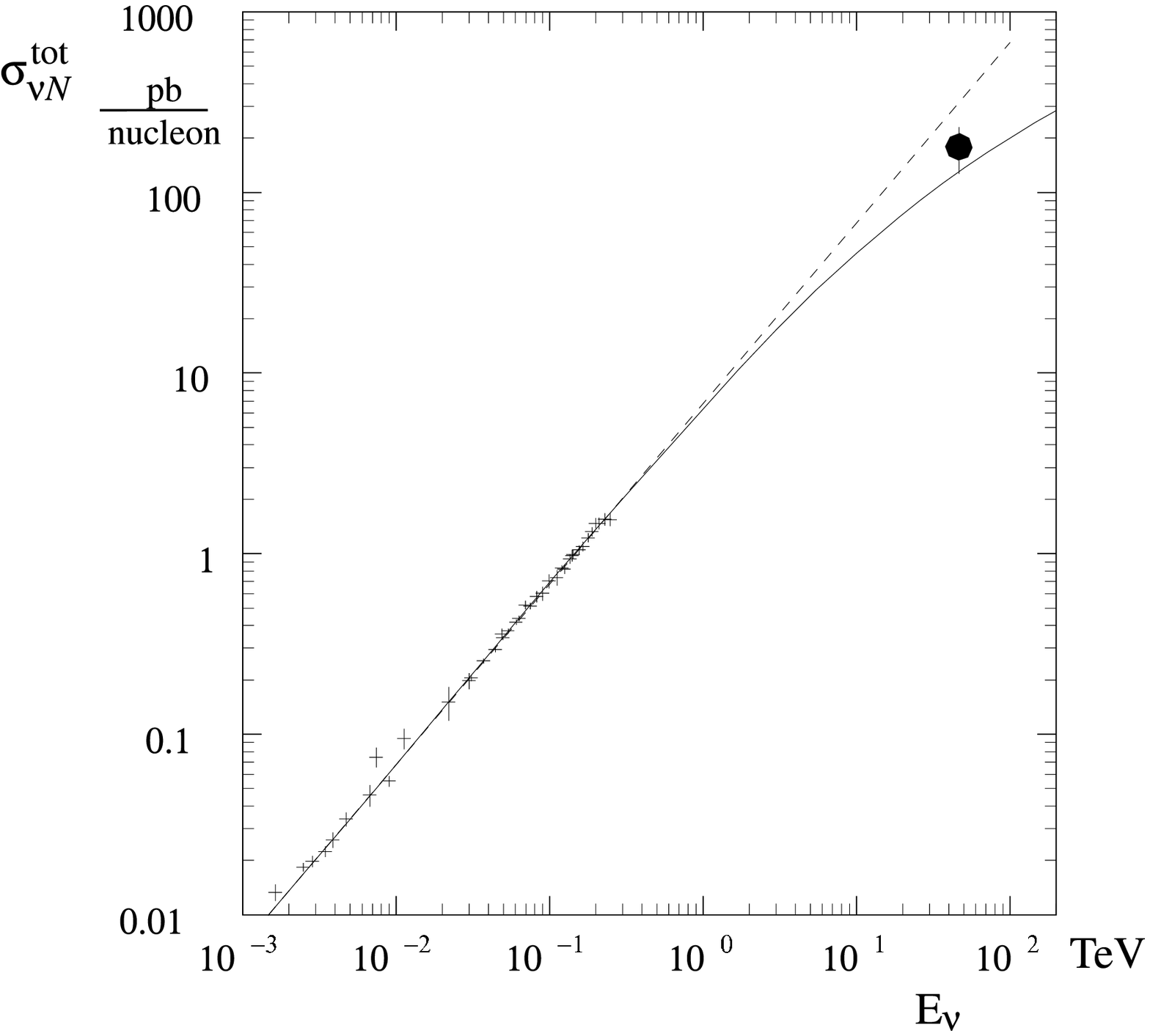}}
\hspace{0.1cm}
\subfigure{\includegraphics[width=2.7in]{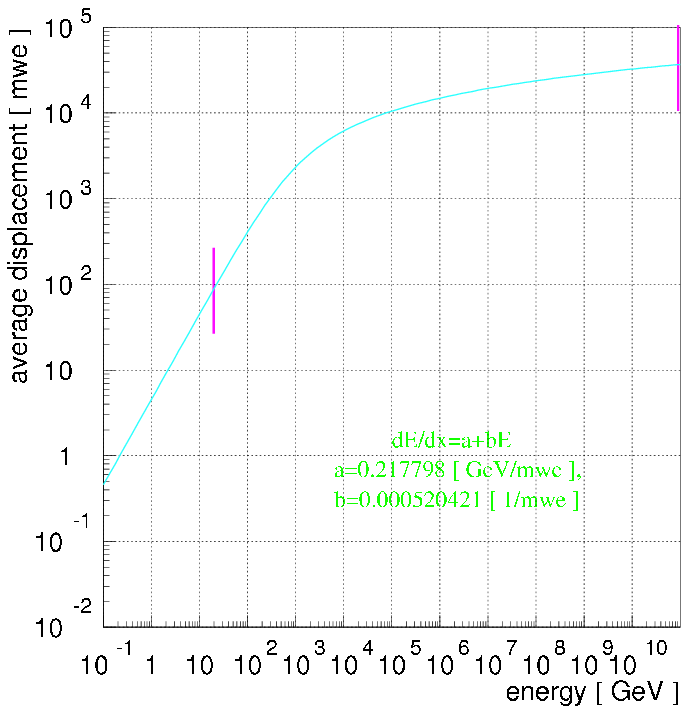}}
\end{center}
\caption{Muon neutrino cross section $\sigma_{\nu}$\cite{HERA} (left) and  muon range $\lambda_{\mu}$ as a function of the neutrino energy\cite{crosssection} (right).}
\label{fig:nu+mu}
\end{figure}
  
Similar arguments apply to the detection of tau neutrinos. A tau neutrino will be detected provided the tau lepton it produces reaches the instrumented volume within its lifetime. Therefore, in Eq.\,4 $L$ is replaced by
\begin{equation}
L \rightarrow \gamma c \tau = E/m c \tau\,,
\end{equation}
where $m$, $\tau$ and $E$ are the mass, lifetime and energy of the tau, respectively. The tau's decay length $\lambda_{\tau} = \gamma c \tau  \approx 50~{\rm m} \times (E_\tau/10^6\,{\rm GeV})$ grows linearly with energy and actually exceeds the range of the muon near 1\,EeV. At yet higher energies the tau eventually ranges out by catastrophic interactions, just like the muon, despite the reduction of the cross sections by a factor $(m_{\mu}/m_{\tau})^2$.

The larger cross sections of neutrinos, the longer range of the muon and the longer lifetime of the tau at high energies make the construction of neutrino detectors of kilometer-scale dimension possible above a threshold of $\sim 100$\,GeV. Muons and tau neutrinos can be detected over volumes of ice and water larger than those instrumented with PMTs; see Fig.\,6a. In Fig.\,6b we show the effective volume for electromagnetic showers of IceCube illustrating that also the  ``effective" volume exceeds the 1\,km$^3$ volume instrumented.

\begin{figure}[t]
\centering\leavevmode
\subfigure[]
{\includegraphics[width=3.0in]{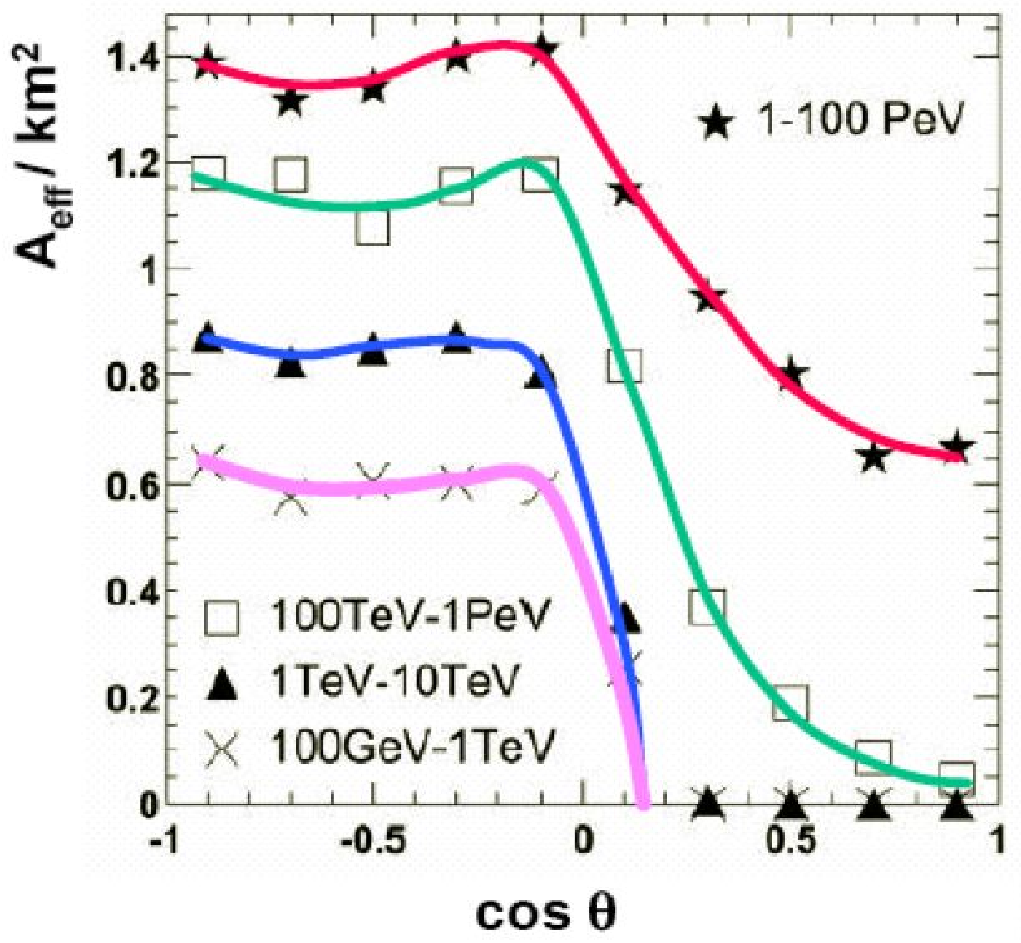}}
\quad
\subfigure[]
{\includegraphics[width=2.7in]{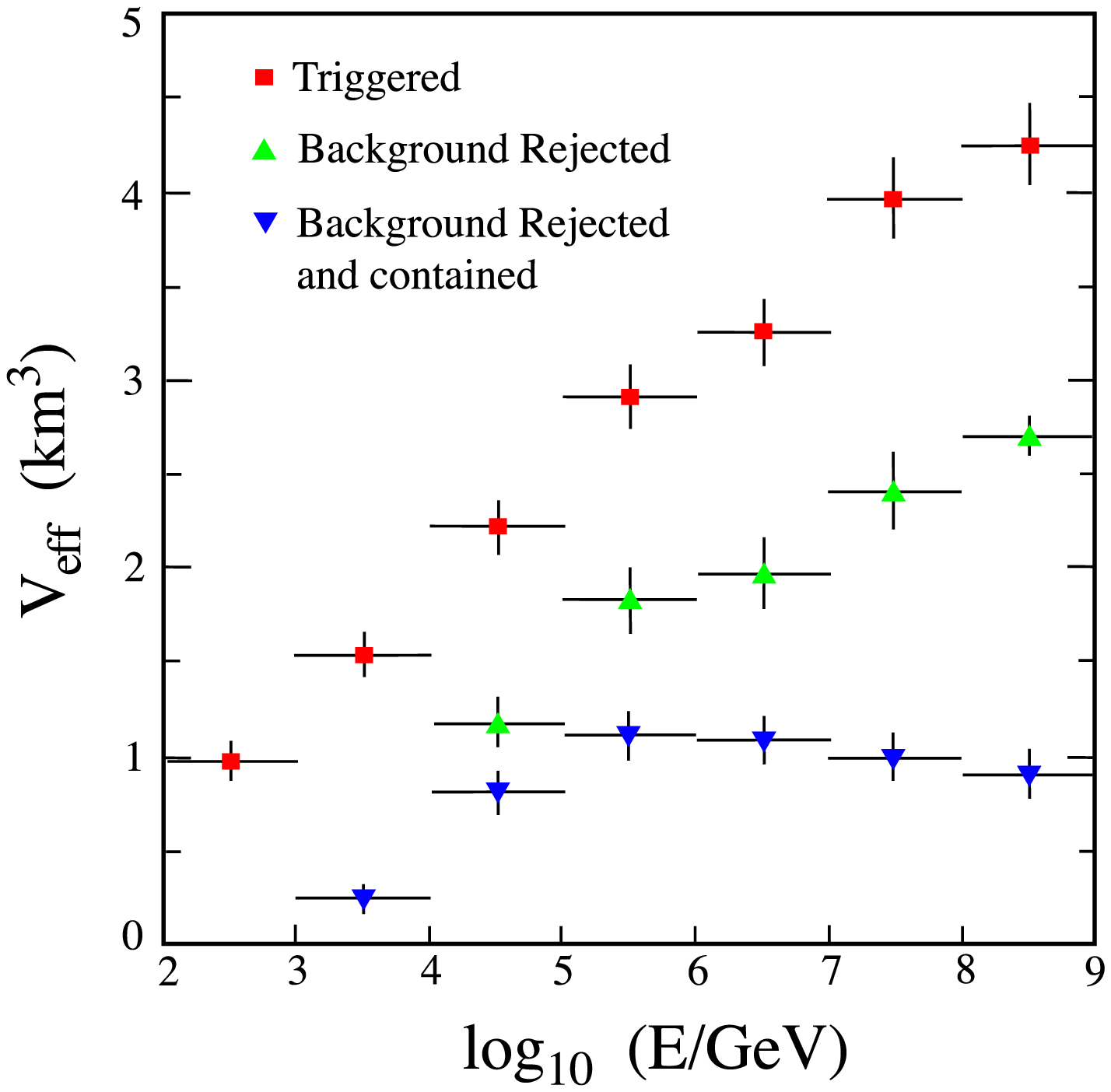}}
	
\caption{Fig.\,6a shows the effective area of IceCube for the detection of neutrinos of muon flavor. Fig.\,6b shows the effective volume of IceCube for the detection of showers initiated by neutrinos of electron or tau flavor.  (The performance of IceCube has been simulated with AMANDA analog signals rather than with the superior digital signals of IceCube and its performance is therefore expected to be superior to what is shown\cite{ice3}.)}
\end{figure}

\subsection{Identification of Neutrino Flavors}

Neutrino telescopes detect the Cherenkov light radiated by secondary particle showers produced by neutrinos of all flavors. These include the electromagnetic and hadronic showers initiated by $\nu_e$ and $\nu_\tau$ as well as by neutral current interactions of neutrinos of all flavors. Because the size of these showers, of order 10\,m in ice, is small compared to the spacing of the PMTs, they represent, to a good approximation, a point source of Cherenkov photons radiated by the shower particles. These trigger the PMTs at the single photoelectron level over a spherical volume whose radius scales linearly with the shower energy; see Fig.\,7.

Whereas the smaller first-generation telescopes mostly exploit the large range of the muon to increase their effective area for $\nu_{\mu}$, kilometer-scale detectors can fully exploit the advantages associated with the detection of showers initiated by $\nu_e$ and $\nu_{\tau}$:
\begin{enumerate}
\item They are detected over both Northern and Southern hemispheres. (We should note that this is also the case for $\nu_{\mu}$ with energy in excess of 1\,PeV where the background from the steeply falling atmospheric spectrum becomes negligible.) IceCube's sensitivity to the galactic center is similar to that of ANTARES, although not to that of a kilometer-scale detector in the Northern hemisphere\cite{volkas}.
\item  The background of atmospheric neutrinos is significantly reduced. At higher energies the muons from $\pi$ decay, the source of atmospheric $\nu_{e}$, no longer decay and relatively rare K-decays become the dominant source of background electron neutrinos.
\item Their energy measurement is superior.
\item $\nu_{\tau}$ are not absorbed, but degraded by energy in the earth.
\end{enumerate}

The detection of neutrinos of all flavors has become especially important for two reasons: neutrino oscillations and tau neutrino ``regeneration" in the earth. The generic cosmic accelerator produces neutrinos from the decay of pions with admixture $\nu_e:\nu_\mu:\nu_\tau = 1:2:0$.  This is also the admixture expected in the atmospheric neutrino beam below 10 GeV where the muons decay. Because of neutrino oscillations the ratio detected is modified to $1:1:1$ because approximately one half of the muon neutrinos convert to tau flavor over cosmic baselines.  This represents an advantage because $\nu_\tau$, unlike $\nu_e$ and $\nu_\mu$, are not absorbed in the Earth. The reason is simple\cite{H&S}. A $\nu_{\tau}$ interacting in the Earth will produce a secondary $\nu_{\tau}$ of lower energy, either directly in a neutral current interaction or via the decay of a tau lepton produced in a charged current interaction. High energy $\nu_{\tau}$ will thus cascade down to PeV energy where the Earth is transparent. In other words, they are detected with a reduced energy but not absorbed.

\subsubsection{Electron Neutrinos} 

Depending on energy, electron neutrinos deposit 0.5-0.8\% of their energy into an electromagnetic shower initiated by the leading final state electron. The rest of the energy goes into the fragments of the target that produce a second subdominant shower. For ice, the Cherenkov light generated by shower particles spreads over a volume of radius 130\,m at 10\,TeV and 460\,m at 10\,EeV, i.e. the shower radius grows by just over 50\,m per decade in energy. 

\begin{figure}[h]
\centering\leavevmode
\includegraphics[width=5in]{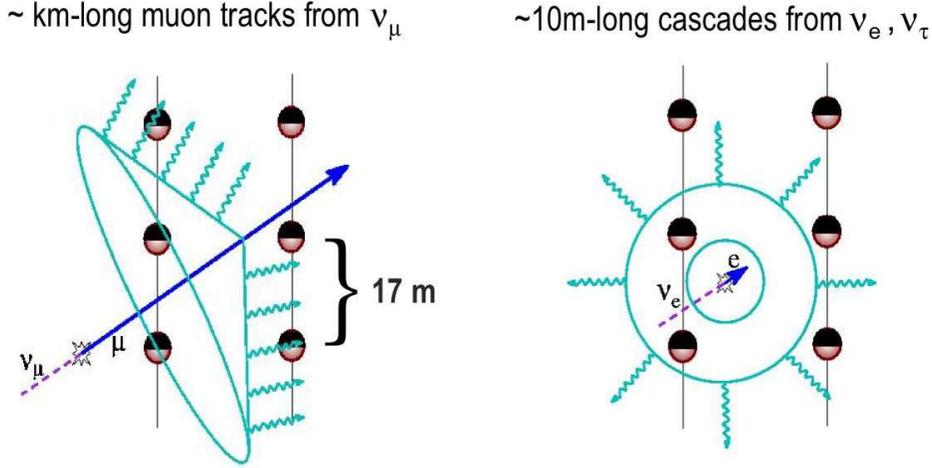}
\caption{Contrasting Cherenkov light patterns produced by muons (left) and by secondary showers initiated by electron and tau neutrinos (right).}
\end{figure}

The measurement of the radius of the lightpool mapped by the lattice of PMTs determines the energy and turns neutrino telescopes into total absorption calorimeters. Note that even a contained ``direct hit" by a 10 EeV neutrino will not saturate a km$^3$ detector volume. So, even for GZK neutrinos, IceCube will not saturate and their energy spectrum can be measured. 

Because the shower and its accompanying Cherenkov lightpool are not totally symmetric but elongated in the direction of the leading electron, the direction of the incident neutrino can be reconstructed. Pointing is however inferior to what can be achieved for muon neutrinos and estimated to be precise to ${\sim}10$\ degrees only. The reconstruction is expected to be better, of order a few degrees, for a significant fraction of the events -- this is a work in progress.

\subsubsection{Muon Neutrinos}

Secondary muons initiated by muon neutrinos range out over kilometers at TeV energy to tens of kilometers at EeV energy, generating showers along their track by bremsstrahlung, pair production and photonuclear interactions; recall Fig.\,5. These are the sources of Cherenkov radiation and are detected in exactly the same way as the leading electron neutrino in the previous section. Because the energy of the muon degrades along its track, also the energy of the secondary showers diminishes and the distance from the track over which the associated Cherenkov light can trigger a PMT is gradually reduced. The geometry of the lightpool surrounding the muon track is therefore a kilometer-long cone with gradually decreasing radius. At the lower energies, hundreds of GeV and less, the muon becomes minimum ionizing. One can perform approximate simulations of IceCube using this approach\cite{halzenhooper}.

High energy muons lose energy catastrophically according to
\begin{equation}
\frac{dE}{dX}=-\alpha - \beta E \, ,
\end{equation}
where $\alpha=2.0 \times 10^{-6}~{\rm TeV}\, {\rm cm}^2/{\rm g}$ and $\beta=4.2 \times
10^{-6}~{\rm cm}^2/{\rm g}$. The distance a muon travels before its energy drops below some energy threshold, $E^{\rm th}_\mu$, called the muon range is then given by
\begin{equation}  
\lambda_\mu = \frac{1}{\beta} \ln \left[ 
\frac{\alpha + \beta E_\mu}{\alpha + \beta E^{\rm th}_\mu} \right] \,.
\label{murange}
\end{equation}

In the first kilometer a high energy muon typically loses energy in a couple of showers of one tenth its initial energy. So the initial size of the cone is the radius of a shower with 10\% of the muon energy, e.g.\ 130\,m for a 100\,TeV muon. Near the end of its range the muon becomes minimum ionizing emitting light that creates single photoelectron signals at a distance of just over 10\,m from the track. For 0.3 photoelectrons, the standard PMT threshold setting, this distance reaches 45\,m; see Fig.\,8.

\begin{figure}[h]
\centering\leavevmode
\includegraphics[width=4in]{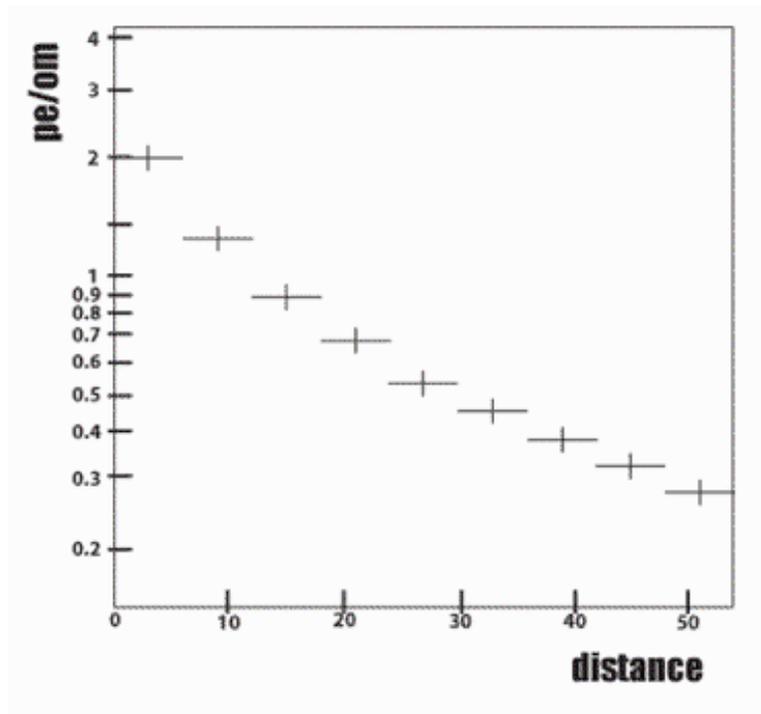}
\caption{Distance in meters over which a 10 inch photomultiplier (om), set at a photoelectron threshold (pe), detects a minimum ionizing muon in ice. }
\end{figure}

Note however that, unlike for showers, the energy measurement is indirect. Because of the stochastic nature of muon energyloss, the logarithm of the energy is measured. Also, although at PeV energy and above, muons have ranges of tens of kilometers, greatly enhancing their detectability, the initial energy of the event cannot always be measured. A muon can be produced at one energy, travel several kilometers, and be detected with much less energy.

\subsubsection{Tau Neutrinos} 

Because half of the muon neutrinos convert over cosmic distances to tau neutrinos whose flux is not attenuated by the earth, their detection has become a priority. Production of $\nu_\tau$ in the beam dump is suppressed relative to $\nu_e$ and $\nu_\mu$ by some five orders of magnitude.  In the absence of oscillations, $\nu_\tau$ of astrophysical origin would have been undetectable. With oscillations they become 1/3 of the cosmic beam and have the additional advantage not to be absorbed by the Earth-- they may loose energy but always reach the detector.

Whereas at lower energies $\nu_\tau$ produce showers indistinguishable from those initiated by $\nu_e$,  the flavor of tau neutrinos of sufficiently high energy can be identified.  Perhaps the most striking signature is the double bang event\cite{LP} in which the production and decay of a $\tau$ lepton are detected as two separated showers inside the detector. It may
also be possible to identify ``lollipop'' events in which a $\nu_\tau$ creates a long minimum-ionizing track that penetrates the detector and ends in a high energy cascade when the $\tau$ lepton decays.  The parent $\tau$ track can be identified by the reduced catastrophic energy loss compared to a muon of similar energy. In other words, the large energy of the shower observed is not compatible with the radiation pattern of a muon; a muon would have revealed its flavor by abundant radiation along the initial track.

The efficiency for a kilometer-scale detector to identify double-bang events can be estimated as follows. In a charged current interaction of a $\nu_\tau$ with a nucleus, a $\tau$ lepton of energy $(1-y)E_{\nu_\tau}$ is produced in association with a hadronic shower of energy $y E_{\nu_\tau}$ from the fragmentation of the target. Here $y$ is the fraction of energy transferred to the hadronic vertex in the interaction. Before decaying, the $\tau$ lepton travels on average a distance $\lambda_\tau$ given by:
\begin{equation}
\lambda_\tau={E_\tau\over m_\tau} ct_0={(1-y)E_{\nu_\tau}\over m_\tau}c\tau
\label{taurange}
\end{equation}
where $E_\tau$ and $m_\tau$ are the energy and mass of the $\tau$ respectively and $\tau$ is its lifetime at rest.  The decay produces another $\nu_\tau$ and an electromagnetic or hadronic shower $\sim82\%$ of the times.  Assuming a detector of dimension $L$, there are several conditions that have to be fulfilled for  identification of a double bang event:

\begin{itemize}

\item The $\nu_\tau$ has to interact through a charged current interaction producing a hadronic shower contained inside or close to the instrumented volume ($\sim L^3$).

\item The $\tau$ lepton must decay inside the detector to a final state
that produces an electromagnetic or hadronic shower which also has to
be contained.

\item $\lambda_\tau$ has to be sufficiently large for the two showers to be clearly separated.

\item The showers must be sufficiently energetic to trigger the detector.

\end{itemize}

In the vicinity of 10\,PeV the probability to detect and identify a $\nu_\tau$ as a double-bang is only 10\% of that for detecting a $\nu_\mu$ of the same energy. At lower and higher energies the likelihood of detecting a double-bang falls rapidly.

\section{High Energy Neutrino Telescopes: Status}

In this section we discuss the first first-generation detector AMANDA with effective telescope area of $1\sim8\times10^4m^2$, depending on the science. We will briefly mention the efforts to build a similar detector in the Mediterranean and discuss the relative performance of water and ice as a Cherenkov medium. This much debated question can now be answered by comparing AMANDA's performance with detailed simulations of the ANTARES detector.

\subsection{AMANDA: First ``Light"}

While it has been realized for many decades that the case for neutrino astronomy is compelling, the challenge has been to develop a reliable, expandable and affordable detector technology to build the kilometer-scale telescopes required to do the science.

Conceptually, the technique is simple. The AMANDA detector, using natural 1 mile-deep Antarctic ice as a Cerenkov detector, has been operated for more than 5 years in its final configuration of 667 optical modules on 19 strings; see Fig.\,9. The detector is in steady operation collecting roughly $7 \sim 10$ neutrinos per day using fast on-line analysis software. At the lower rate a background-free sample is obtained all the way to the horizon. The challenge has been to detect these neutrinos in the presence of a background of down-going cosmic ray muons that trigger the detector at a rate of $\sim$80\,Hz, or a signal to background ratio of order one million; see Fig.\,10. The rejection is achieved by reconstruction and angular cuts. AMANDA's performance has been calibrated by reconstructing atmospheric muons as well as muons produced by atmospheric muon neutrinos\cite{pune}.

\begin{figure}[t]
\centering\leavevmode
\includegraphics[height=5in]{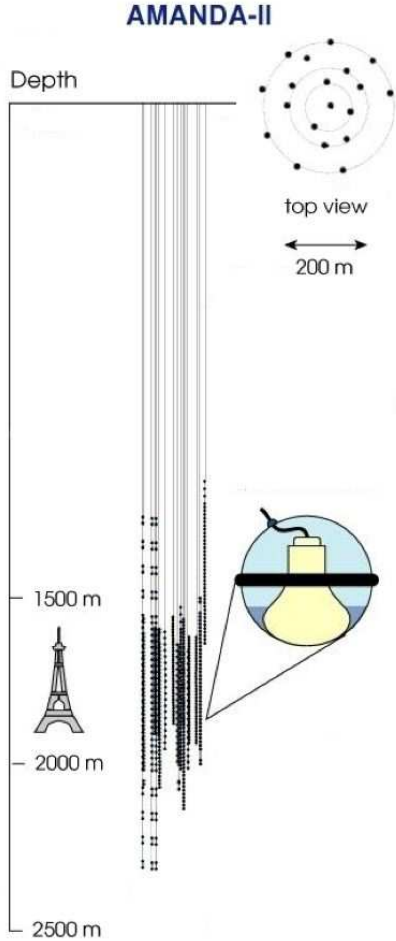}
\caption{AMANDA instruments $1.6 \times 10^7$\,m$^3$ of ice with 677 photomultipliers deployed on 19 strings. }
\end{figure}

\begin{figure}[t]
\subfigure[]{\includegraphics[width=2.4in]{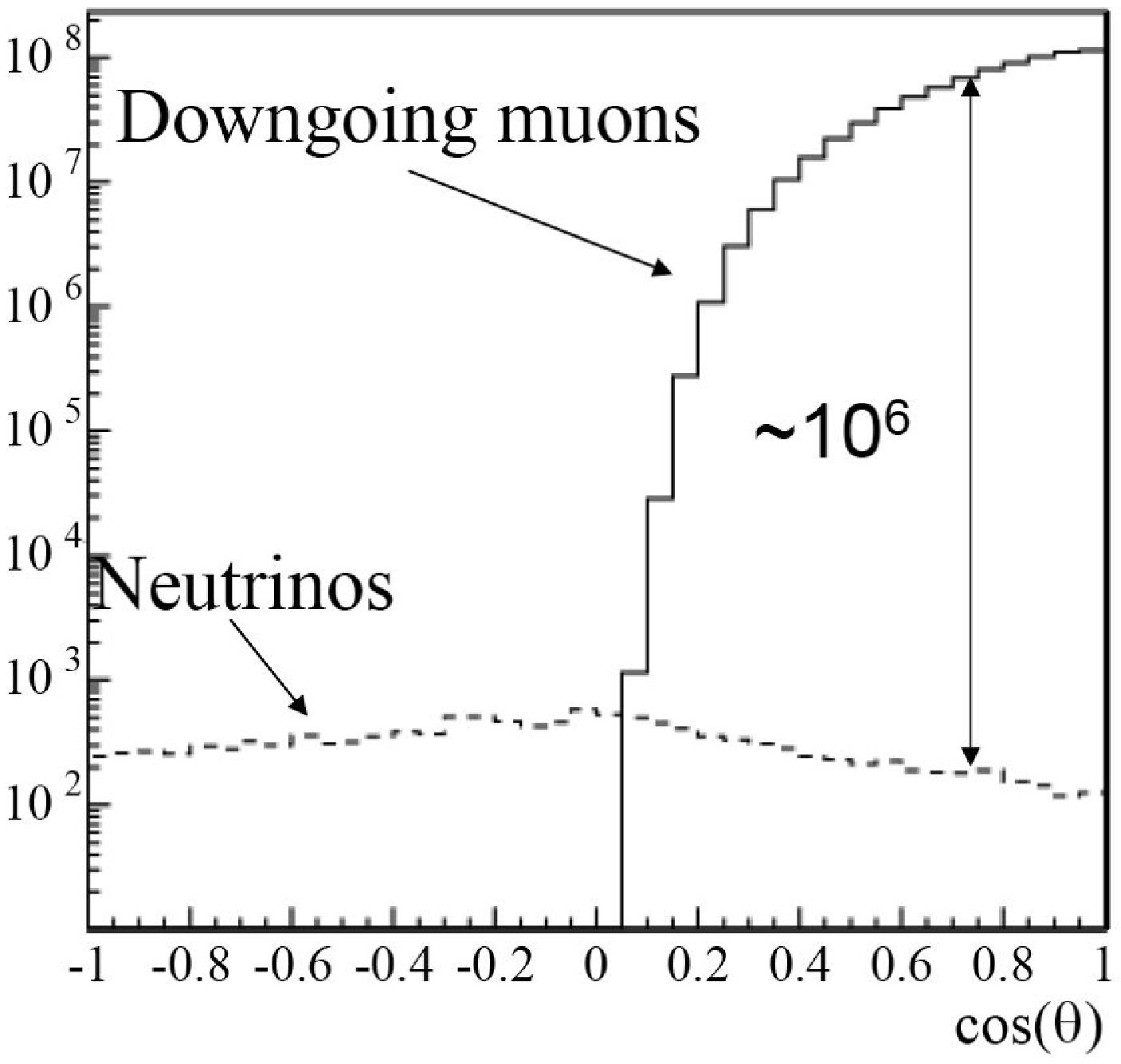}}
\quad
\subfigure[]{\includegraphics[width=3.4in]{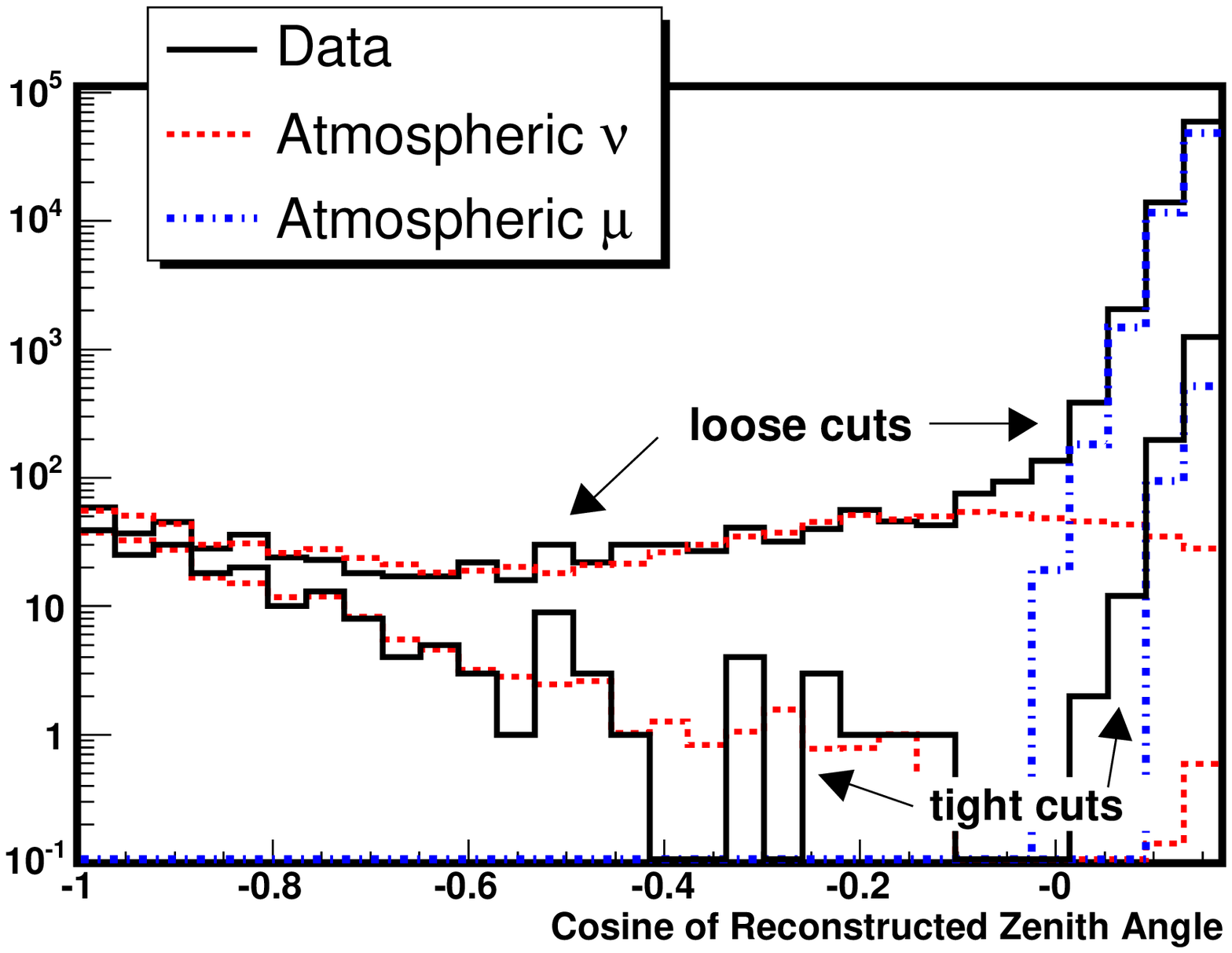}}
\caption{(a) Flux of upward moving atmospheric neutrino-induced muons and downward moving atmospheric muons as a function of zenith angle. (b) As the quality selection of the events is tightened, a clear separation between the two classes of events is achieved in the AMANDA data.}
\end{figure}

\begin{figure}[h]
\centering\leavevmode
\includegraphics[width=4.25in]{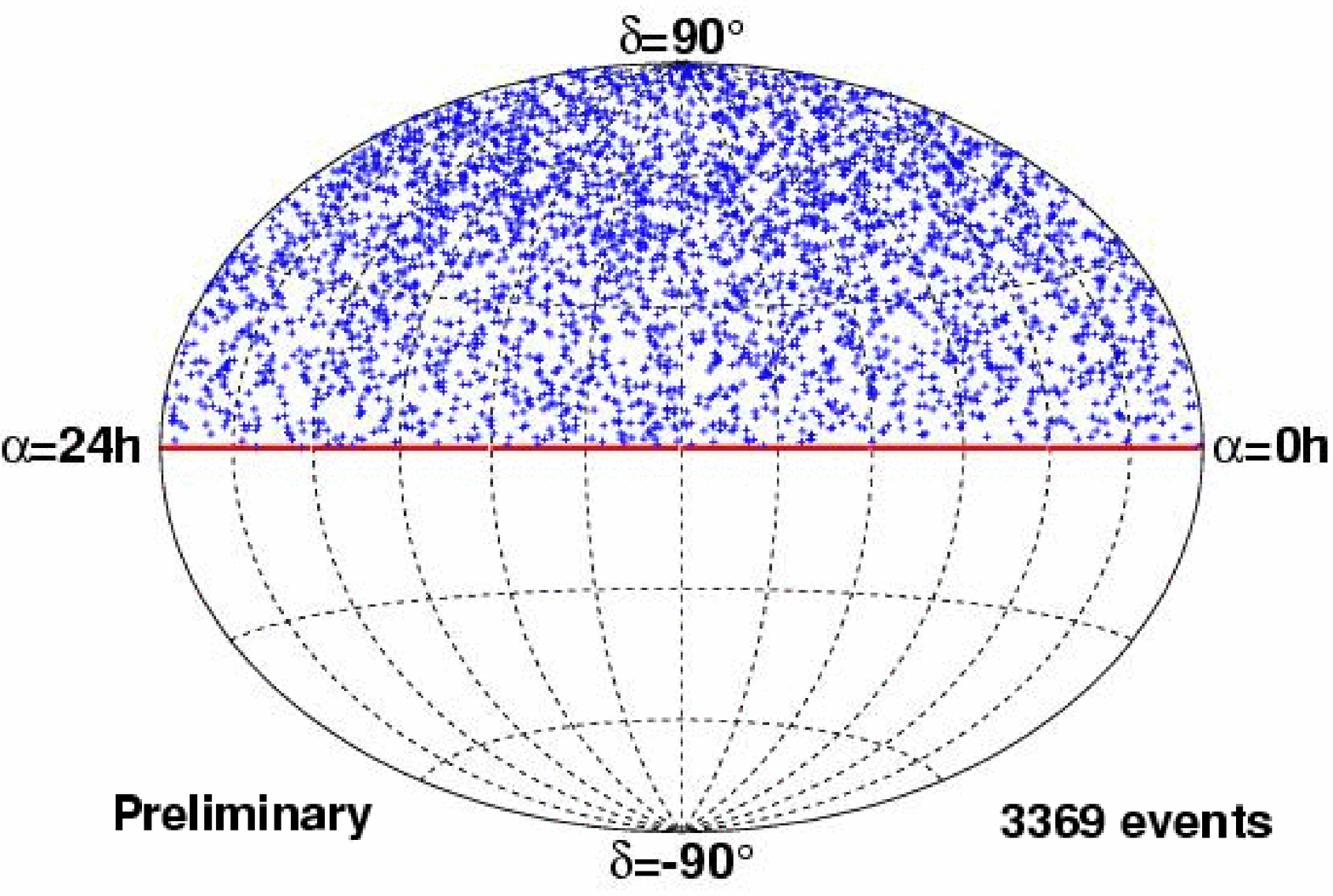}
\caption[]{Skymap showing the declination and right ascension of neutrinos detected by the AMANDA\,II detector during four Antarctic winters of operation in 2000-2003.\label{fig:skyplot}}
\end{figure}

Using the first 4 years of AMANDA data, the collaboration is performing a search for  the emission of muon neutrinos from spatially localized directions in the northern sky\cite{pune,HS}. The neutrino arrival directions are shown for 800 days of data in a skyplot of declination and right ascension; see Fig.~\ref{fig:skyplot}. The 90\% upper limits on the neutrino fluency of point sources are at the level of $6 \times 10^{-8}\rm\, GeV \,cm^{-2}\, s^{-1}$ or $10^{-10}\rm \,erg\ cm^{-2}\,s^{-1}$, averaged over declination. This corresponds to a flux of $6 \times 10^{-9}\rm\, cm^{-2}\, s^{-1}$ integrated above 10\,GeV assuming a $E^{-2}$ energy spectrum typical for shock acceleration of particles in non-thermal high energy sources. The most significant excess is 3.4\,$\sigma$ from the Crab with a probability of close to 10\% given the trial factor for 33 sources searched; see Table~1. IceCube is needed to make conclusive observations of sources.

The AMANDA detector has reached a high-energy effective telescope area of 25,000$\sim$40,000\,m$^2$, depending on declination; see Fig.\,12. This represents an interesting milestone\cite{alvarezhalzen}: known TeV gamma ray sources, such as the active galaxies Markarian 501 and 421, should be observed in neutrinos if the number of gamma rays and neutrinos emitted are roughly equal as expected from cosmic ray accelerators producing pions. Therefore AMANDA must detect the observed TeV photon sources soon, or, its observations will exclude them as significant sources of cosmic rays.

\begin{figure}[t]
\centering\leavevmode
\includegraphics[width=4.25in]{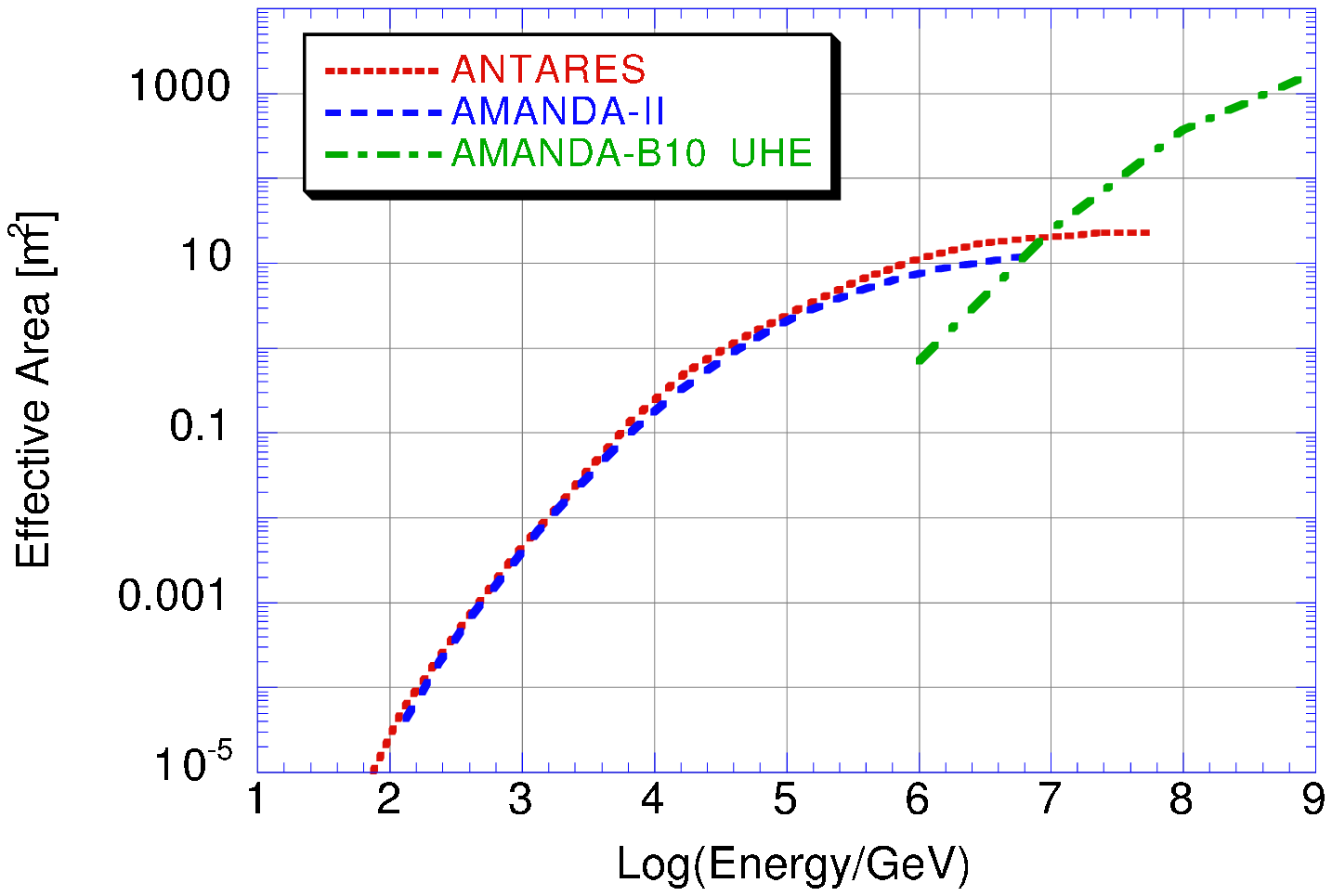}
\caption{Effective area of the ANTARES and Amanda detectors as a function of energy. Shown separately is the effective area of the AMANDA detector (10 strings only) for ultra high energy events.}
\end{figure}

\begin{table}[t]
  \caption{\label{ps-table}
                Results from the AMANDA-II search for neutrinos from selected objects.
                $\delta$ is the declination in degrees, $\alpha$ the
                right ascension in hours,
                $n_{obs}$ is the number of observed events, and $n_{b}$
                the expected
                background. $\Phi_{\nu}^{\mathrm{lim}}$ is the
                90\% CL upper limit in units of $10^{-8}
                \mathrm{cm}^{-2}\mathrm{s}^{-1}$
                for a spectral index of 2
                and integrated above 10 GeV.
                These results are preliminary (systematic errors
                are not included).}
 \begin{center}
 \footnotesize
 \tabcolsep0.4em
  \begin{tabular}{lccccc|lccccc}\hline\hline
      Candidate & $\delta$($^\circ$) & $\alpha$(h)    &
                $n_{\mathrm{obs}}$ & $n_{b}$          &
                $\Phi_{\nu}^{\mathrm{lim}}$           &
      Candidate & $\delta$($^\circ$) & $\alpha$(h)    &
                $n_{\mathrm{obs}}$ & $n_{b}$          &
                $\Phi_{\nu}^{\mathrm{lim}}$           \\\hline
\multicolumn{12}{c}{ \emph{TeV Blazars} } \\
      Markarian 421  & 38.2 & 11.07 & 6 & 5.6 & 0.68&
      1ES 2344+514   & 51.7 & 23.78 & 3 & 4.9 & 0.38\\
      Markarian 501  & 39.8 & 16.90 & 5 & 5.0 & 0.61&
      1ES 1959+650   & 65.1 & 20.00 & 5 & 3.7 & 1.0 \\
      1ES 1426+428   & 42.7 & 14.48 & 4 & 4.3 & 0.54&
                     &&&&&\\
\multicolumn{12}{c}{ \emph{GeV Blazars} } \\
      QSO 0528+134   & 13.4 &  5.52 & 4 & 5.0 & 0.39&
      QSO 0219+428   & 42.9 &  2.38 & 4 & 4.3 & 0.54\\
      QSO 0235+164   & 16.6 &  2.62 & 6 & 5.0 & 0.70&
      QSO 0954+556   & 55.0 &  9.87 & 2 & 5.2 & 0.22\\
      QSO 1611+343   & 34.4 & 16.24 & 5 & 5.2 & 0.56&
      QSO 0716+714   & 71.3 &  7.36 & 1 & 3.3 & 0.30\\
      QSO 1633+382   & 38.2 & 16.59 & 4 & 5.6 & 0.37&
                     &&&&&\\
  \multicolumn{12}{c}{ \emph{Microquasars} } \\
      SS433          &  5.0 & 19.20 & 2 & 4.5 & 0.21&
      Cygnus X3      & 41.0 & 20.54 & 6 & 5.0 & 0.77\\
      GRS 1915+105   & 10.9 & 19.25 & 6 & 4.8 & 0.71&
      XTE J1118+480  & 48.0 & 11.30 & 2 & 5.4 & 0.20\\
      GRO J0422+32   & 32.9 &  4.36 & 5 & 5.1 & 0.59&
      CI Cam         & 56.0 &  4.33 & 5 & 5.1 & 0.66\\
      Cygnus X1      & 35.2 & 19.97 & 4 & 5.2 & 0.40&
      LS I +61 303   & 61.2 &  2.68 & 3 & 3.7 & 0.60\\
   \multicolumn{12}{c}{ \emph{SNR \& Pulsars} }\\
      SGR 1900+14    &  9.3 & 19.12 & 3 & 4.3 & 0.35&
      Crab Nebula    & 22.0 &  5.58 &10 & 5.4 & 1.3\\
      Geminga        & 17.9 &  6.57 & 3 & 5.2 & 0.29&
      Cassiopeia A   & 58.8 & 23.39 & 4 & 4.6 & 0.57\\
  \multicolumn{12}{c}{ \emph{Miscellaneous} }\\
      3EG J0450+1105 & 11.4 &  4.82 & 6 & 4.7 & 0.72&
      J2032+4131     & 41.5 & 20.54 & 6 & 5.3 & 0.74\\
      M 87           & 12.4 & 12.51 & 4 & 4.9 & 0.39&
      NGC 1275       & 41.5 &  3.33 & 4 & 5.3 & 0.41\\
      UHE CR Doublet & 20.4 &  1.28 & 3 & 5.1 & 0.30&
      UHE CR Triplet & 56.9 & 11.32 & 6 & 4.7 & 0.95\\
      AO 0535+26     & 26.3 &  5.65 & 5 & 5.0 & 0.57&
      PSR J0205+6449 & 64.8 &  2.09 & 1 & 3.7 & 0.24\\
      PSR 1951+32    & 32.9 & 19.88 & 2 & 5.1 & 0.21&
                     &&&&&\\
       \hline\hline
       \end{tabular}
 \end{center}
  \end{table}

The sensitivity of the detector can be enhanced by leveraging special properties of the sources. A partial list of possibilities includes:
\begin{itemize}
\item Whereas we previously limited the discussion of the sensitivity of AMANDA to the "Waxman-Bahcall" flux to muon neutrinos (see Fig.\,3), limits on a diffuse flux of neutrinos of all flavors were also established. These are reported in Table\,2 for the assumption of a $E^{-2}$ cosmic beam with flavor composition $\nu_e$:$\nu_{\mu}$:$\nu_{\tau}$ of 1:1:1.  
\item Of special interest in Table\,2 is the limit for very high energy events, PeV and above. Because the earth is essentially opaque to neutrinos with energies in excess of $\sim10$\,PeV, cosmic neutrinos can only penetrate to the detector near the horizon. Separation of such a signal from bundels of down-going atmospheric muons near the horizon can be performed by energy measurement. Extremely energetic signal events produce a higher light density compared to the lower energy atmospheric muon bundles that represent the dominant background. Therefore signal events can be selected by the multiple photon signals they produce in a large number of PMTs. This analysis results in an increased effective area in the highest energy range, even for the 300\,PMT detector operated in 97-99; see Fig.\,12.
\item For GRB one can limit the search to a window in temporal and directional coincidence the with satellite observations. This dramatically reduces the background and results in an effective area that can reach as high as $0.08$\,km$^2$. The background is directly determined by off-source events.
\item The GRB approach can be applied to time-variable sources such as flaring AGN. For the blazar 1ES 1959+650 3 of 5 events observed in a four year period cluster in an interval of 66 days overlapping with a period of very high activity of the source in TeV gamma rays. One of the events is coincident with an ``orphan" flare, i.e. without X-ray counterpart suggesting hadronic origin of the TeV gamma rays\cite{pune}.
\item In order to observe sources producing signals at the level of the sensitivity limit of the telescope, several categories of selected sources were ``stacked" and tested for a cumulative signal.
\item The passage of a large flux of MeV-energy neutrinos during a period of seconds will be detected as an excess of the background counting rate in all individual optical modules\cite{hjz}. The AMANDA detector monitors most of the galaxy, including the galactic center, and is part of the SNEWS network\cite{snews}. IceCube will reach all the way to the Large Magellanic Cloud and will yield a high statistics measurement of the time evolution of the source. IceCube has the potential to do neutrino physics using a MeV energy supernova neutrino beam. This includes the possibility of measuring the third angle in the neutrino mixing matrix by coincident observation of the supernova with a Northern hemisphere detector such as SuperK. Sensitivity to $\theta_{13}$ results from matter effects on the beam traveling through the earth\cite{raffelt}.
\end{itemize}
Updated reports on AMANDA results can be found in reference\cite{pune}. It is tempting and inevitable to speculate on possible signals of weak statistical significance in the present data. It is much more useful to build a more sensitive detector, especially as all robust estimates indicate the necessity.

\begin{table}
\caption{{\bf Summary of AMANDA diffuse neutrino flux results, 1997-2003}.
The results labeled ``muon'' are for analyses sensitive to neutrino-induced
muon tracks in the detector, and give limits on the muon-neutrino flux at
earth. The ``all-flavour'' analyses are sensitive to
events from muon, electron and tau neutrinos, and place limits on the 
total neutrino flux at the earth, assuming a 1:1:1 flavour ratio  due
to maximal mixing neutrino oscillations during propagation to the earth. 
Assuming this 1:1:1 flavour ratio, 
the muon-neutrino limits may be
 converted to all-flavour limits
 by multiplying by three.}
\label{diffusetable}
  \centering\leavevmode
  \footnotesize
    \begin{tabular}{l@{\hspace{1em}}c@{\hspace{.2em}}c@{\hspace{-1em}}c@{\hspace{-.8em}}c}
      \hline
      {\bf Data set} & {\bf Detection channel} & {\bf $\mathbf{\nu}$ energy range} & {Limit \esqdnde ($\mathrm{90\%c.l.}$)} & {Reference}  \\
       & &{\bf TeV} & {\bf \diffunit} & \\
      \hline \hline
 1997 & muon & $6-10^3$  & $8.4 \times 10^{-7}$ & \cite{HS} \\
 1997 & all flavour & $10^3-3\times10^6$ & $9.9 \times 10^{-7}$ & \cite{1997UHE} \\
 1997 & all flavour & $50-3\times10^3$ & $9.8 \times 10^{-6}$ & \cite{nature} \\
 2000 & all flavour  & $50-5\times10^3$   & $8.6 \times 10^{-7}$ & \cite{2000cascade} \\
 2000 & all flavour  &$1.8\times10^2-1.8\times10^6$& \llap*$3.8\times10^{-7}$ & \cite{pune} \\
 2000 & muon, unfolding & $100-300$ & $2.6 \times 10^{-7}$ & \cite{pune} \\
2000-03 & muon & $16-2\times10^3$ & \llap*$8.9 \times 10^{-8}$ & \cite{pune} \\
\hline \hline
    & & & *analysis in progress, sensitivity only \\
       \hline
\end{tabular}
\end{table}

\subsection{Mediterranean Telescopes}

Below PeV energy, South Pole telescopes are not sensitive to a flux of $\nu_\mu$ from sources in the Southern sky, which is obscured by the large flux of cosmic ray muons. This and the obvious need for more than one telescope --- accelerator experiments have clearly demonstrated the value of multiple detectors --- provide compelling arguments for deploying northern detectors. With the first observation of neutrinos by a detector in Lake Baikal with a telescope area of 2500\,m$^2$ for TeV muons\cite{baikal} and after extensive R\&D efforts by both the ANTARES\cite{ANTARES} and NESTOR\cite{nestor} collaborations in the Mediterranean, there is optimism that the technological challenges to build neutrino telescopes in deep sea water have been met.  Both Mediterranean collaborations have demonstrated their capability to deploy and retrieve optical sensors, and have reconstructed down-going muons with optical modules deployed for R\&D tests.

The ANTARES neutrino telescope is under construction at a 2400\,m deep Mediterranean site off Toulon, France. It will consist of 12 strings, each equipped with 75 optical sensors mounted in 25 triplets. The detector performance has been fully simulated\cite{ANTARES} with the following results: a sensitivity after one year to point sources of $0.4-5 \times 10^{-7}\rm\, GeV\, cm^{-2}\, s^{-1}$ and to a diffuse flux of $0.9 \times 10^{-7}\rm\, GeV \,cm^{-2}\, s^{-1}$ above 50\,TeV. As usual, a $E^{-2}$ spectrum has been assumed for the signal\cite{karle}.

It is interesting to contrast these simulation result with the performance of AMANDA because it gives us insight into the complex question of the relative merits of water and ice as a Cherenkov medium. The comparison is relatively straightforward because AMANDA and ANTARES operate at similar depths.  With 600 8-inch modules used in a typical data analysis versus 900 10-inch photomultipliers with 35\% larger photocathode assumed in ANTARES simulations, AMANDA is a factor of two smaller. We will see that ANTARES roughly matches the sensitivity of 800 days of AMANDA data. The conclusion seems to be that the telescope sensitivity is the same for equal photocathode area.

The relative performance of water and ice has been debated without reaching consensus because the comparison is not simple:
\begin{enumerate}
\addtolength{\itemsep}{-.5ex}
\item Ice absorbs less with absorption lengths exceeding 100\,m even at the dominant blue wavelengths of Cherenkov light where the absorption length in water is only of order 10\,m.
\item Water scatters less with scattering lengths of hundreds of meters. Depending on depth and the color the scattering length in ice can be as low as a few meters and only exceeds 50\,m in the last 350\,m instrumented by IceCube.
\item Background counting rates of the PMT are 40\,kHz or more in water and less than 1\,kHz in ice.
\end{enumerate} 

In four Antarctic winters, or about 800 days of data, AMANDA has reached similar point source limits\cite{HS} of $0.6 \times 10^{-7}\rm\, GeV\, cm^{-2}\, s^{-1}\,sr^{-1}$ as ANTARES simulations anticipate in one year of data taking. Also the diffuse limits reached in the absence of a signal are comparable\cite{pune}. We have summarized the sensitivity of both experiments in Table~3 where they are also compared to the sensitivity of IceCube.

\begin{table}[t]
\caption{Sensitivity of IceCube, AMANDA, and ANTARES to point and diffuse sources of cosmic neutrinos. Tabulated is the flux $E_\nu^2 \, dN/dE_\nu$ in units of Gev cm$^{-2}$\, s$^{-1}$\, (sr$^{-1}$).}
\centering\leavevmode
\small
\def\arraystretch{1.33}
\begin{tabular}{@{}|c|c|c|c|}
\hline
& \bf  IceCube& \bf AMANDA-II$^*$& \bf ANTARES\\
\hline
\bf \# of PMTs& 4800 / 10 inch& 600 / 8 inch& 900 / 10 inch\\
\hline
\parbox{7em}{{ \bf Point source\\ sensitivity}\\ (neutrinos/year)}& $5 \times 10^{-9}$&
\parbox{8.5em}{\baselineskip11pt\centering $1.3 \times 10^{-7}$ weakly dependent\\ on declination}&
\parbox{9.5em}{\baselineskip11pt\centering 0.4--$5\times 10^{-7}$ depending\\ on declination}\\[.2in]
\hline
\parbox{7em}{{\bf diffuse limit$^\dagger$}\\ (neutrinos/year)}& 
\parbox{7.5em}{\centering 3--$12\times 10^{-9}$}&
\parbox{5.5em}{\centering $2\times 10^{-7}\rm$}&
\parbox{6.5em}{\centering $0.9\times 10^{-7}$}\\[.1in]
\hline
\multicolumn{4}{l}{$^*$includes systematic errors}\\[-1ex]
\multicolumn{4}{l}{$^\dagger $depends on assumption for background from atmospheric neutrinos from charm}
\end{tabular}
\end{table}

The superior angular resolution of ANTARES ($< 0.5^\circ$) compared to AMANDA ($1.8^\circ$) does translate into a better sensitivity to point sources. The average upper limits are compared in Fig.\,13 which shows similar results, though reached in 800 days of AMANDA data taking compared to one year of ANTARES Monte Carlo. The result can be understood as follows\cite{teresa}. The average upper limit is proportional to the square root of the number of (atmospheric neutrino) background events in the search bin around the source which is 0.5 and 2 degrees for ANTARES and AMANDA, respectively, with the limit set at only 3 compared to 5 events.

\begin{figure}[h!]
\centering\leavevmode
\includegraphics[width=3in]{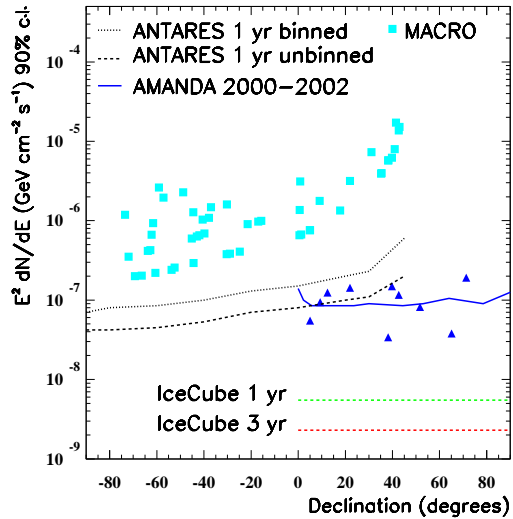}
\caption{Flux limits on point sources of cosmic neutrinos derived from Macro (squares, 6 years) and AMANDA (triangles, ${\sim}800$ days) data. Also shown are future upper limits anticipated for ANTARES and IceCube after 1 year of operation.}
\end{figure}

In the same context, the NEMO collaboration has done the interesting exercise of simulating the IceCube detector (augmented from 4800 to 5600 optical modules; see next section) in water rather than ice. One finds a slightly reduced sensitivity in water, probably not significant within errors and at no energy larger than 50\%\cite{emigneco}.

\subsection{Kilometer-scale Neutrino Observatories}

The sensitivities in Table 2  imply that in several years of operation a kilometer-scale detector like IceCube can improve the sensitivity of first-generation telescopes by two orders of magnitude. The baseline design of kilometer-scale neutrino detectors maximizes sensitivity to $\nu_\mu$-induced muons with energy above hundreds of GeV, where the acceptance is enhanced by the increasing neutrino cross section and muon range and the Earth is still largely transparent to neutrinos. The mean-free path of a $\nu_\mu$ becomes smaller than the diameter of the Earth above 70\,TeV --- above this energy neutrinos can only reach the detector from angles closer to the horizon. Good identification of other neutrino flavors becomes a priority, especially because they provide complete angular coverage and because $\nu_\tau$ are not absorbed by the Earth. Good angular resolution is required to distinguish possible point sources from background, while energy resolution is needed to enhance the signal from astrophysical sources, which are expected to have flatter energy spectra than the background atmospheric neutrinos.

Overall, AMANDA represents a proof of concept for the kilometer-scale neutrino observatory, IceCube\cite{ice3}, now under construction. IceCube will consist of 80 kilometer-length strings, each instrumented with 60 10-inch photomultipliers spaced by 17~m. The deepest module is 2.4~km below the surface. The strings are
arranged at the apexes of equilateral triangles 125\,m on a side. The instrumented detector volume is a cubic kilometer. A surface air shower detector, IceTop, consisting of 160 Auger-style 2.7\,m diameter ice-filled Cherenkov detectors deployed over 1\,km$^{2}$ above IceCube, augments the deep-ice component by providing a tool for calibration, background rejection and cosmic ray studies; see Fig.\,14. Where cosmic ray physics is concerned, IceCube represents an excellent opportunity to study the spectrum and composition in the key energy range near and above the ``knee".

\begin{figure}[!h]
\centering\leavevmode
\subfigure[]{\includegraphics[height=3.25in]{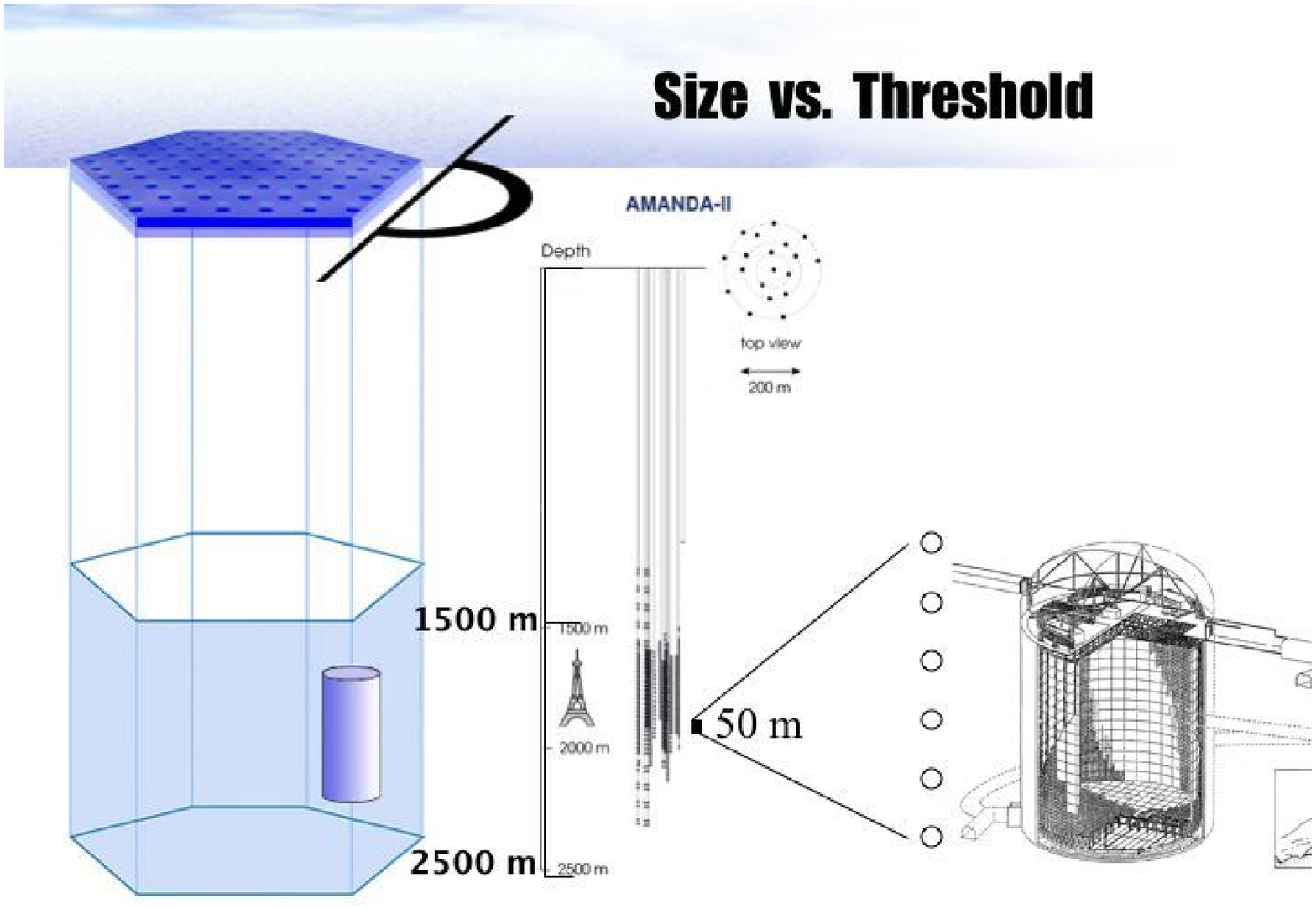}}
\quad
\subfigure[]{\includegraphics[height=3.25in]{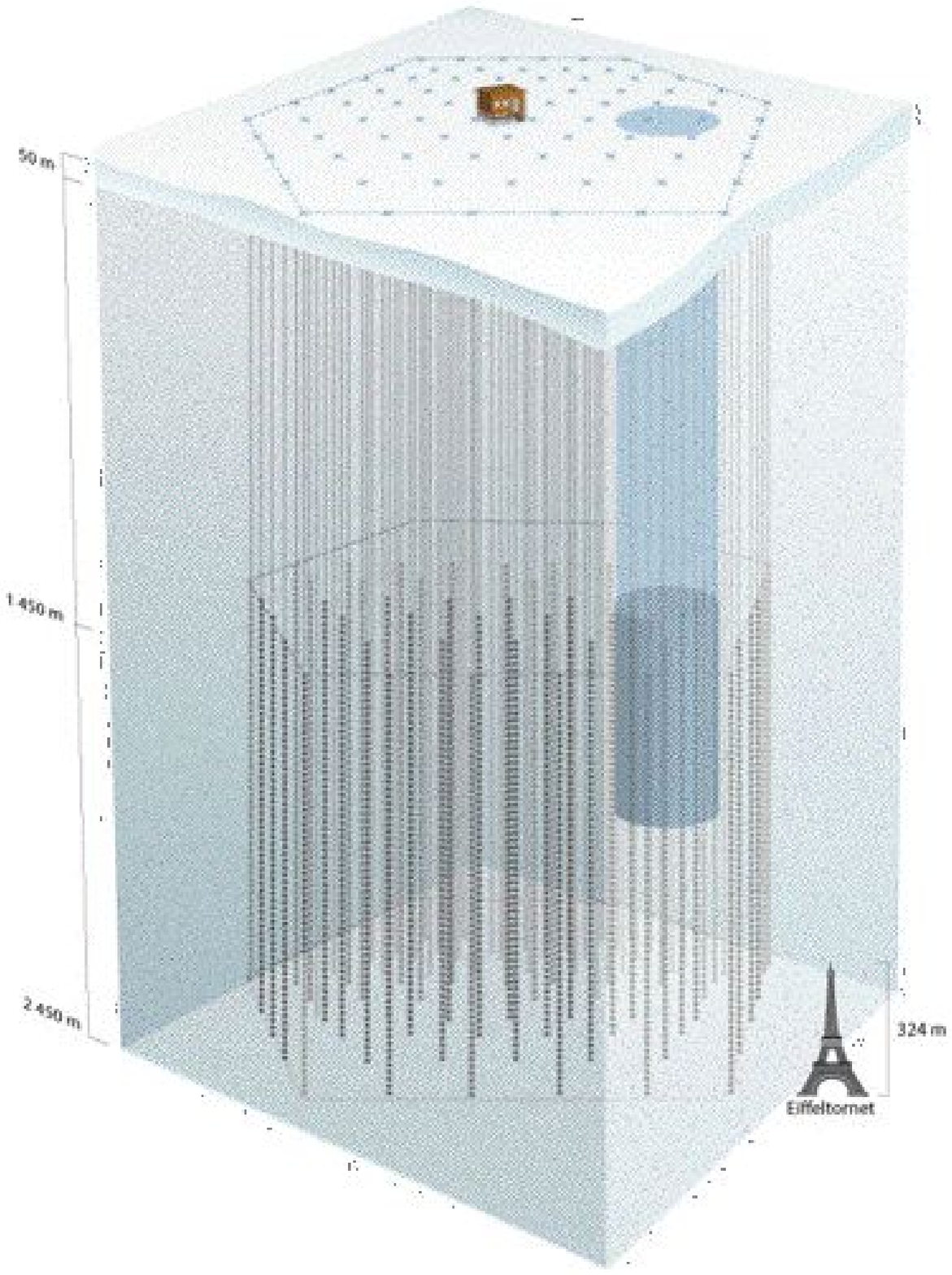}}
\caption{(a) Relative sizes of the IceCube, AMANDA, and Superkamiokande neutrino detectors. AMANDA will be operated as a lower threshold subsystem of IceCube. As the size of the detector grows, so does the threshold energy of neutrinos detected. (b) Architecture of IceCube.}
\end{figure}

The transmission of analog photomultiplier signals from the deep ice to the surface, used in AMANDA, has been abandoned. The photomultiplier signals will be captured and digitized inside the optical module to minimize the loss of information from degradation of the signals sent over long cables.  The digitized signals are given a global time stamp with residuals less than $3$\,ns and transmitted to the surface.  The digital messages are sent to a string processor, a global event trigger and an event builder.

Each digital optical module functions independently. The PMT output is collected with custom waveform-digitizer chips that sample the signal 128 times at 200 to 700 megasamples per second. The PMT signal is fed into 3 parallel 10 bit digitizers with a nominal gain ratio 0.25:2:16. Combine they provide a dynamic range of 14 bits of resolution, covering single photoelectrons to complete PMT saturation. Late-arriving light is recorded with a 40 MHz, 10 bit analog-to-digital converter that stores 256 samples over 6.4 microseconds. A large programmable gate array with an embedded processor controls the system and compresses and packages the data. A block diagram of the IceCube main board is shown in Fig.\,15. The entire digital module operates on 5\,W of power.

Construction of the detector commenced in the Austral summer of 2004/2005 with the assembly of the the 5 megawatt hot water drill that should drill to 2500\,m in less than 2 days. At the end of the season a hole was delivered in 52 hours and the first 60 digital optical modules deployed in about 20 hours.\footnote{In the 05--06 Antarctic summer another 8 strings were deployed. At this point IceCube already exceeds AMANDA in light collection area.} The first 8 IceTop tanks have been also deployed. Neutrino events have been observed. The growing detector will take data during construction, with each string coming online within days of deployment. All modules performed as designed and timing of 2\,ns has been demonstrated over the full string. With minimal deadtime, all modules collect supernova signals at a background counting rate of 300\,Hz on average. In this low background environment, IceCube can detect the excess of MeV anti-$\nu_e$ events from a supernova out to the Magellanic clouds.

The data streams of IceCube, and AMANDA, embedded inside IceCube,  will be merged. The present schedule calls for completion in 2010, although a km$^2$\,year of data will be acquired as soon as 2007.

\begin{figure}[h]
\centering\leavevmode
\includegraphics[width=6in]{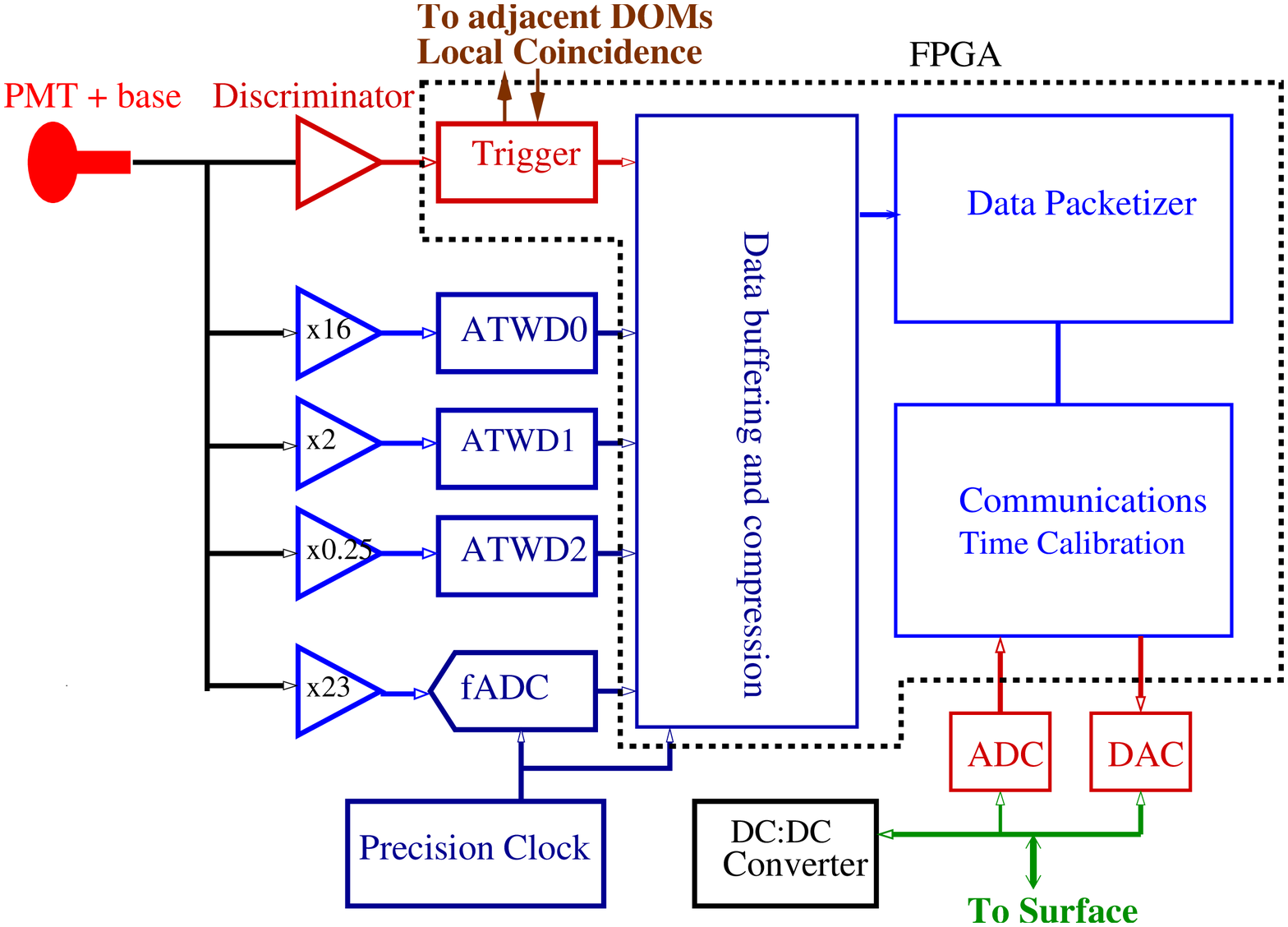}
\caption{Block diagram of an IceCube main board.}
\end{figure}

IceCube will offer advantages over AMANDA beyond its larger size: it will have a higher efficiency and superior angular resolution, map showers from electron- and tau-neutrinos and, most importantly, measure neutrino energy. Simulations, benchmarked by AMANDA data, indicate that the direction of muons can be determined with sub-degree accuracy and
their energy measured to better than 30\% in the logarithm of the energy. The direction of showers will be reconstructed to better than 10$^\circ$ above 10\,TeV and the response in energy is linear and better than 20\%. Energy resolution is critical because, once one establishes that the energy exceeds 1\,PeV, there is no atmospheric muon or neutrino background in a kilometer-square detector and full sky coverage of the telescope is achieved. Samples of simulated events are shown in Fig.\,16.

\begin{figure}[h]
\centering\leavevmode
\includegraphics[width=4in]{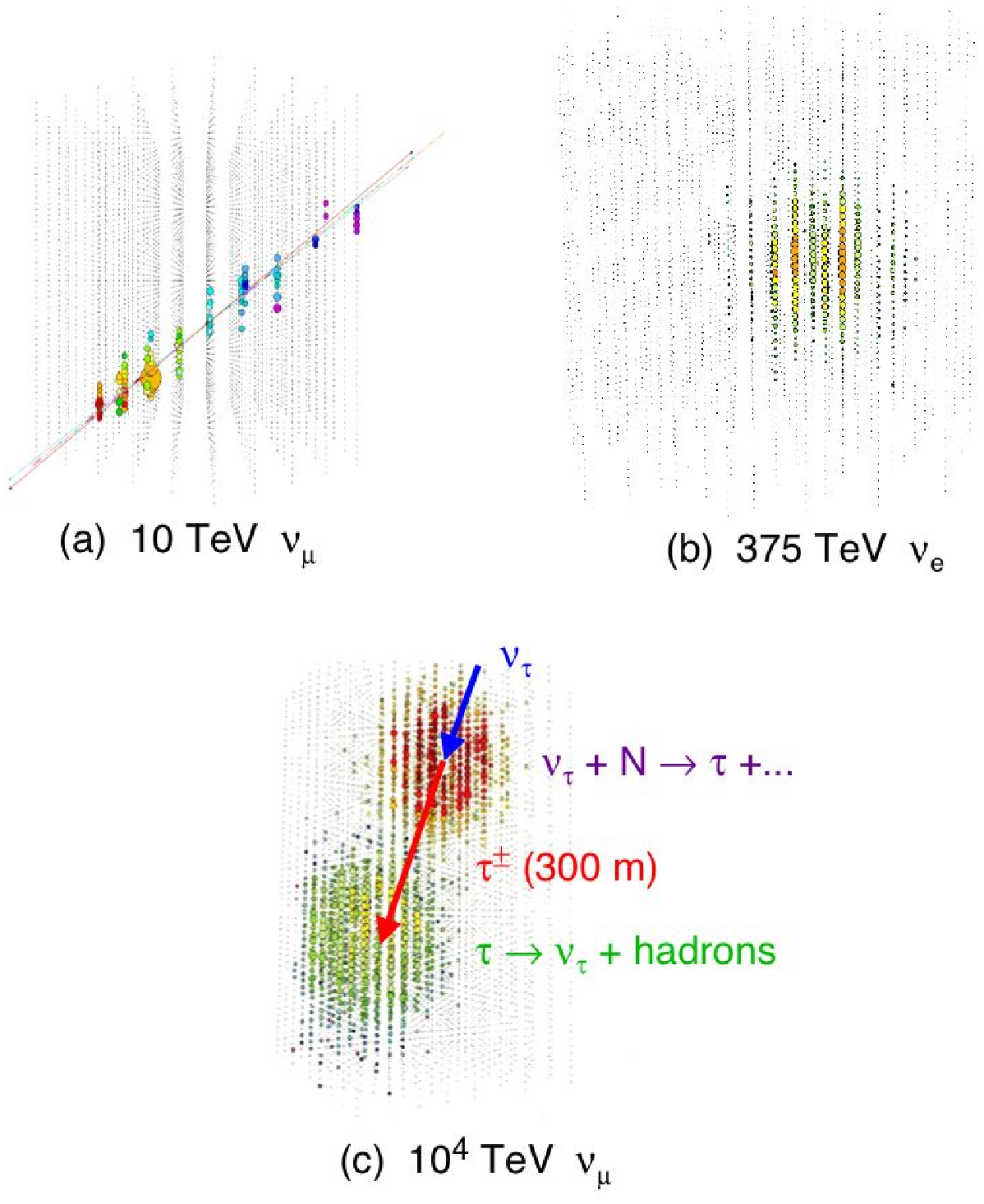}
\caption{Simulation of IceCube events: (a) a 10\,TeV $\nu_{\mu}$, (b) a 375\,TeV $\nu_e$ and (c) a $10^4$\,TeV $\nu_{\tau}$. The secondary tau decays after 300\,m.}
\end{figure}

NEMO, an INFN R\&D project in Italy, has been mapping Mediterranean sites and studying novel mechanical structures, data transfer systems as well as low power electronics, with the goal to deploy a next-generation detector similar to IceCube. A concept has been developed with 81 strings spaced by 140\,m. Each consists of 18 bars that are 20\,m long and spaced by 40\,m. A bar holds a pair of photomultipliers at each end, one looking down and one horizontally. As already mentioned, the simulated performance\cite{NEMO} is, not unexpectedly, similar to that of IceCube with a similar total photocathode area as the NEMO concept.

Recently, a wide array of projects have been initiated to detect neutrinos of the highest energies, typically above a threshold of 10 EeV, exploring other experimental signatures: horizontal air showers and acoustic or radio emission from neutrino-induced showers. Some of these experiments, such as the Radio Ice Cerenkov Experiment\cite{frichter} and an acoustic array in the Caribbean\cite{lehtinen}, have taken data; others are under construction, such as the Antarctic Impulsive Transient Antenna\cite{gorham}. 

\section*{Acknowledgments}
I thank my IceCube collaborators and Luiz Anchordoqui, Concha Gonzalez-Garcia, Gary Hill, Dan Hooper, Albrecht Karle and Teresa Montaruli for discussions. This research was supported in part by the National Science Foundation under Grant No.~OPP-0236449, in part by the U.S.~Department of Energy under Grant No.~DE-FG02-95ER40896, and in part by the University of Wisconsin Research Committee with funds granted by the Wisconsin Alumni Research Foundation.

\section{Appendix: Formalism for calculating the\\ number of neutrinos detected from a source\\ producing a neutrino flux \boldmath{$\phi(E_\nu)$}}

For ``realistic" calculations of the number of events detected from a source producing a neutrino flux $\phi(E_\nu)$, a more complete formalism is required. We will compile the necessary formulae here.

\subsection{Muon Tracks from $\nu_{\mu}$}

The complete expression to compute the expected number of $\nu_\mu$ induced events is given by
\begin{eqnarray}
N^{\nu_\mu}_{\rm ev}
&=& T \int^{1}_{-1} d\cos\theta\,  
\int^\infty_{l'min} dl\,
\int_{m_\mu}^\infty dE_\mu^{\rm fin}\,
\int_{E_\mu^{\rm fin}}^\infty dE_\mu^0\, 
\int_{E_\mu^0}^\infty dE_\nu \\ \nonumber
&&\frac{d\phi_{\nu_\mu}}{dE_\nu d\cos\theta}(E_\nu,\cos\theta)
\frac{d\sigma^\mu_{CC}}{dE_\mu^0}(E_\nu,E_\mu^0)\, n_T\, 
F(E^0_\mu,E_\mu^{\rm fin},l)\, A^0_{eff}\, \, .
\label{eq:numuevents}
\end{eqnarray}
$\frac{d\phi_{\nu_\mu}}{dE_\nu d\cos\theta}$ is the differential $\nu_\mu$ flux in the vicinity of the detector after propagation through the Earth matter. Because of the high energies of the neutrinos their oscillations, propagation in the Earth and regeneration by $\tau$ decay must be treated in a coherent way\cite{GHM}. $\frac{d\sigma^\mu_{CC}}{dE_\mu^0}(E_\nu,E_\mu^0)$ is the differential charged current interaction cross section producing a muon of energy $E_\mu^0$ and
$n_T$ is the number density of nucleons in the matter surrounding the
detector and $T$ is the exposure time of the detector. After production with energy $E_\mu^0$, the muon ranges out in the
rock and in the ice surrounding the detector and looses energy.  $F(E^0_\mu,E_\mu^{\rm fin},l)$ the function that describes
the energy spectrum of the muons arriving at the detector.  Thus
$F(E^0_\mu,E_\mu^{\rm fin},l)$ represents the probability that a muon
produced with energy $E_\mu^0$ arrives at the detector with energy
$E_\mu^{\rm fin}$ after traveling a distance $l$. The function $F(E^0_\mu,E_\mu^{\rm fin},l)$ is computed by propagating the muons to
the detector taking into account energy losses due to ionization,
bremsstrahlung, $e^+e^-$ pair production and nuclear interactions; see for instance Ref.~\cite{ls}.

Equivalently, muon events arise from $\bar\nu_\mu$ interactions. They are calculated from an equation similar to Eq.\,15 with appropriate substitutions for antineutrinos.

\subsection{Showers from $\nu_e$ and $\nu_{\tau}$} 

The shower event rate at IceCube can be obtained using the following semi-analytical calculations~\cite{Anchordoqui:2005gj},
\begin{equation}
{N}_{\rm sh}  = 
{N}_{\rm sh,CC}+{N}_{\rm sh,NC} \,\,,
\label{FTjpg}
\end{equation}
where
\begin{equation}
{N}_{\rm sh,CC} =  T\, n_{\rm T}\, 
\int_{E^{\rm min}_{\rm sh}}^\infty {\rm d}E_{\nu} \, \sum_{\alpha=e,\tau}\frac
{d\phi^{\nu_\alpha}}{dE_{\nu}}(E_{\nu})
\sigma_{\rm CC}(E_{\nu})\, {V}_{\rm eff} (E_\nu) \,\,,
\label{eq:shsourcecc}
\end{equation}
and
\begin{equation}
{N}_{\rm sh,NC} =  T\, n_{\rm T}\,
\int_{E_{\nu}-E^{\rm min}_{\rm sh}}^\infty {\rm d}E'_{\nu}
\int_{E^{\rm min}_{\rm sh}}^\infty {\rm d}E_{\nu} \, \sum_{\alpha=e,\mu,\tau}
\frac{d \phi^{\nu_\alpha}}{dE_{\nu}}(E_{\nu})
\frac{d\sigma_{\rm NC}}{dE'_{\nu}}
(E_{\nu}, E'_{\nu})\, {V}_{\rm eff} (E_\nu) \,\,.
\label{eq:shsourcenc}
\end{equation}
Here, $d\sigma_{\rm NC}/dE'_{\nu}$ is the differential NC
interaction cross section producing a secondary neutrino of
energy $E'_{\nu}$, and  ${V}_{\rm eff} (E_\nu)$ is the effective volume shown in Fig.\,6b. In writing Eqs.~({\ref{eq:shsourcecc}})
and (\ref{eq:shsourcenc}) we identify the shower energy with the $\nu_e$ energy, or $E_{\rm
sh}=E_{\nu}$ in a CC interaction, while for NC interactions the shower
energy corresponds to the energy in the hadronic shower $E_{\rm
sh}=E_{\nu}-E'_{\nu}\equiv E_{\nu}\, y,$
where $y$ is the usual inelasticity parameter in deep inelastic 
scattering~\cite{Halzen:1984mc}.

\end{document}